\documentclass{article}
\usepackage{amssymb}
\usepackage{amsfonts}
\usepackage{amsmath}
\usepackage{graphics}
\usepackage{color}

\input{tcilatex}
\begin{document}

\title{New directions for gravity-wave physics via \textquotedblleft
Millikan oil drops\textquotedblright }
\author{Raymond Y. Chiao \and Professor in the School of Natural Sciences
\and and in the School of Engineering \\
University of California, P. O. Box 2039\\
Merced, CA 95344 \and E-mail: rchiao@ucmerced.edu}
\date{April 6aa, 2007 for the Townes Book \emph{Visions of Discovery}}
\maketitle

\begin{abstract}
Pairs of Planck-mass--scale drops of superfluid helium coated by electrons
(i.e., \textquotedblleft Millikan oil drops\textquotedblright ), when
levitated in the presence of strong magnetic fields and at low temperatures,
can be efficient quantum transducers between electromagnetic (EM) and
gravitational (GR) radiation. A Hertz-like experiment, in which EM waves are
converted at the source into GR waves, and then back-converted at the
receiver from GR waves back into EM waves, should be practical to perform.
This would open up observations of the gravity-wave analog of the CMB from
the extremely early Big Bang, and also communications directly through the
interior of the Earth.
\end{abstract}

\bigskip \emph{And God said, \textquotedblleft Let there be
light,\textquotedblright\ and there was light. (Gen. 1:3)}

\section{Introduction}

In this book in honor of my beloved teacher, colleague, and friend for over
four decades, Professor Charles Hard Townes, I would like to take a fresh
look at an old problem concerning which we had many discussions from early
on, going back to the days when I was his graduate student at M.I.T. After a
visit to Joseph Weber's laboratory at the University of Maryland in the
60's, I still can remember his critical remarks concerning the experiments
then being conducted in Weber's lab using large, massive aluminum bars. He
expressed concerns that the numbers which he calculated indicated that it
would be extremely difficult to see any observable effects, and he was
therefore worried that Weber would not be able to see any genuine signal. \
Later, he expressed to me his similar worries also about LIGO, especially in
light of its large scale and expense. \ 

Here I would like to revisit the problem of generating gravitational
radiation, which has many similarities to that of generating electromagnetic
radiation. The famous work of\ Gordon, Zeiger, and Townes on the maser
opened up entirely new directions in coherent electromagnetic wave research
by generating coherent microwaves by means of the principle of stimulated
emission of radiation. Are there new ideas which might stimulate similar
developments that would open up new directions in gravity-wave research? I
would like to explore here situations in which the principle of reciprocity
(i.e., time-reversal symmetry) demands the existence of non-negligible
back-actions of a measuring device upon the gravitational radiation fields
that are being measured in a quantum mechanical context.

The quantum approach taken here is in stark contrast to the classical,
test-particle approaches being taken in contemporary, large-scale
gravity-wave experiments, which are based solely on classical physics. The
back-actions of classical measuring devices such as Weber bars and large
laser interferometers upon the incident gravitational fields that are being
measured, are completely negligible. Hence they can only detect gravity
waves from powerful astronomical sources such as supernovae \cite{MTW}, but
they certainly cannot generate these waves.

Specifically, I would like to explore here the quantum physics of
Planck-mass--scale \textquotedblleft Millikan oil drops\textquotedblright\
consisting of electron-coated superfluid helium drops at milli-Kelvin-scale
temperatures in the presence of Tesla-scale magnetic fields, as a means to
test whether some of these back-action effects really exist or not. I am in
the process of performing some of these experiments with my colleagues at
the new 10th campus of the University of California at Merced in order to
test some of these ideas. These experiments have become practical to perform
because of advances in ultra-low temperature dilution-refrigerator
technology. I will describe some of these experiments below.

\section{Forces of gravity and electricity between two electrons}

Let us first consider, using only classical, Newtonian concepts (which are
valid in the correspondence-principle limit and at large distances
asymptotically, as seen by a distant observer), the forces experienced by
two electrons separated by a distance $r$ in the vacuum. Both the
gravitational and the electrical force obey long-range, inverse-square laws.
Newton's law of gravitation states that%
\begin{equation}
\left\vert F_{G}\right\vert =\frac{Gm_{e}^{2}}{r^{2}}
\label{Newton's-inverse-square-law}
\end{equation}%
where $G$ is Newton's constant and $m_{e}$ is the mass of the electron.
Coulomb's law states that%
\begin{equation}
\left\vert F_{e}\right\vert =\frac{e^{2}}{r^{2}}\text{ }
\label{Coulomb's-law}
\end{equation}%
where $e$ is the charge of the electron (in Gaussian esu units). The
electrical force is repulsive, and the gravitational one attactive.

Taking the ratio of these two forces, one obtains the dimensionless ratio of
coupling constants%
\begin{equation}
\frac{\left\vert F_{G}\right\vert }{\left\vert F_{e}\right\vert }=\frac{%
Gm_{e}^{2}}{e^{2}}\approx 2.4\times 10^{-43}\text{ .}  \label{Gm^2/e^2}
\end{equation}%
The gravitational force is extremely small compared to the electrical force,
and is therefore usually omitted in all treatments of quantum physics.

\section{Gravitational and electromagnetic radiation powers emitted by two
electrons}

The above ratio of the coupling constants $Gm_{e}^{2}/e^{2}$ is also the
ratio of the powers of gravitational (GR) to electromagnetic (EM) radiation
emitted by two electrons separated by a distance $r$ in the vacuum, when
they undergo an acceleration $a$ and are moving with a speed $v$ relative to
each other, as seen by a distant observer.

From the equivalence principle, it follows that dipolar gravitational
radiation does not exist \cite{MTW}. \ Rather, the lowest order radiation
permitted by this principle is quadrupolar, and not dipolar, in nature.
General relativity predicts that the power $P_{GR}^{\text{(quad)}}$\
radiated by a time-varying mass quadrupole tensor $D_{ij}$ of a periodic
system is given by \cite{MTW}\cite{Landau}\cite{Weinberg}%
\begin{equation}
P_{GR}^{\text{(quad)}}=\frac{G}{45c^{5}}\left\langle \dddot{D}%
_{ij}^{2}\right\rangle =\omega ^{6}\frac{G}{45c^{5}}\left\langle
D_{ij}^{2}\right\rangle   \label{triple-dot}
\end{equation}%
where the triple dots over $\dddot{D}_{ij}$ denote the third derivative with
respect to time of the mass quadrupole-moment tensor $D_{ij}$ of the system
(the Einstein summation convention over the spatial indices $(i,j)$ for the
term $\dddot{D}_{ij}^{2}$ is being used here), $\omega $ is the angular
frequency of the periodic motion of the system, and the angular brackets
denote time averaging over one period of the motion.

Applying this formula to the periodic orbital motion of two point masses
with equal mass $m$ moving with a relative instantaneous acceleration whose
magnitude is given by $\left\vert a\right\vert =\omega ^{2}\left\vert
D\right\vert $, where $\left\vert D\right\vert $ is the magnitude of the
relative displacement of these objects, and where the relative instantaneous
speed of the two masses is given by $\left\vert v\right\vert =\omega
\left\vert D\right\vert $ (where $v<<c$), all these quantities being
measured by a distant observer, one finds that Equation (\ref{triple-dot})
can be rewritten as follows:%
\begin{equation}
P_{GR}^{\text{(quad)}}=\kappa \frac{2}{3}\frac{Gm^{2}}{c^{3}}a^{2}\text{
where }\kappa =\frac{2}{15}\frac{v^{2}}{c^{2}}.  \label{modified-Larmor}
\end{equation}%
The frequency dependence of the radiated power predicted by Equation (\ref%
{modified-Larmor}) scales as $v^{2}a^{2}\sim \omega ^{6}$, in agreement with
triple dot term $\dddot{D}_{ij}^{2}$ in Equation (\ref{triple-dot}). It
should be stressed that the values of the quantities $a$ and $v$ are those
being measured by an observer at infinity. The validity of Equations (\ref%
{triple-dot}) and (\ref{modified-Larmor}) has been verified by observations
of the orbital decay of the binary pulsar PSR 1913+16 \cite{Taylor1994}.

Now consider the radiation emitted by two electrons undergoing an
acceleration $a$ relative to each other with a relative speed $v$, as
observed by an observer at infinity. For example, these two electrons could
be attached to the two ends of a massless, rigid rod rotating around the
center of mass of the system like a dumbbell. The power in gravitational
radiation that they will emit is given by%
\begin{equation}
P_{GR}^{\text{(quad)}}=\kappa \frac{2}{3}\frac{Gm_{e}^{2}}{c^{3}}a^{2}\text{,%
}  \label{GR-Larmor}
\end{equation}%
where the factor $\kappa $ is given above in Equation (\ref{modified-Larmor}%
). Due to their bilateral symmetry, these two identical electrons will also
radiate quadrupolar, but not dipolar, electromagnetic radiation with a power
given by%
\begin{equation}
P_{EM}^{\text{(quad)}}=\kappa \frac{2}{3}\frac{e^{2}}{c^{3}}a^{2}\text{ ,}
\label{EM-Larmor-quad}
\end{equation}%
with the same factor of $\kappa $. The reason that this is true is that any
given electron carries with it mass as well as charge as it moves, since its
charge and mass must co-move rigidly together. Therefore two electrons
undergoing an acceleration $a$ relative to each other with a relative speed $%
v$ will emit simultaneously both electromagnetic and gravitational
radiation, and the quadrupolar electromagnetic radiation which it emits will
be completely homologous to the quadrupolar gravitational radiation which it
also emits.

It follows that the ratio of gravitational to electromagnetic radiation
powers emitted by the two-electron system is given by the same ratio of
coupling constants as that for the force of gravity relative to the force of
electricity, viz.,%
\begin{equation}
\frac{P_{GR}^{\text{(quad)}}}{P_{EM}^{\text{(quad)}}}=\frac{Gm_{e}^{2}}{e^{2}%
}\approx 2.4\times 10^{-43}\text{ .}  \label{Ratio-of-powers}
\end{equation}%
Thus it would seem at first sight to be hopeless to try and use any
two-electron system as the means for coupling between electromagnetic and
gravitational radiation.

Nevertheless it must be emphasized here that although this dimensionless
ratio of coupling constants is extremely small, the gravitational radiation
emitted from the two electron system must \emph{in principle} exist, or else
there must be something fundamentally wrong with the experimentally
well-tested inverse-square laws given by Equations (\ref%
{Newton's-inverse-square-law}) and (\ref{Coulomb's-law}).

\section{The Planck mass scale}

\begin{figure}[tbp]
\centerline{\includegraphics{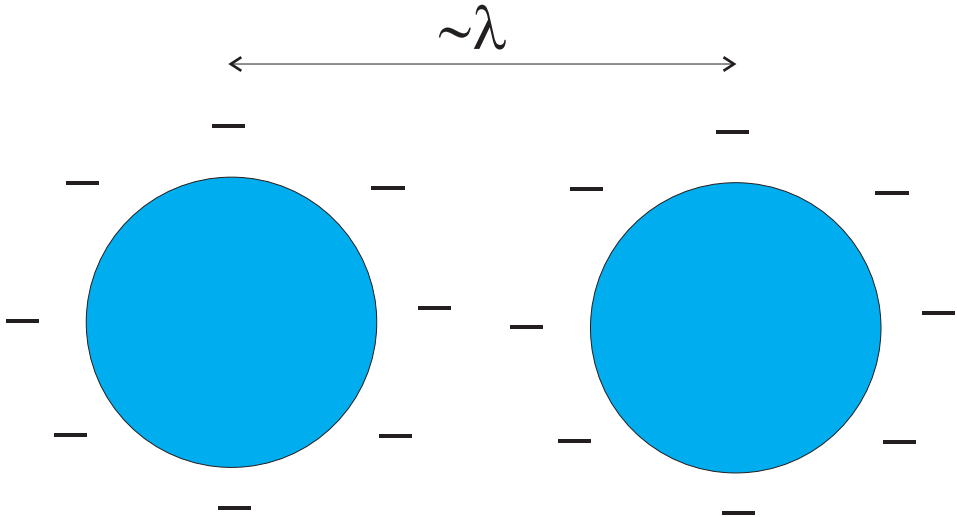}}
\caption{Planck-mass-scale superfluid helium drops coated with electrons on
their outside surfaces and separated by around a microwave wavelength $%
\protect\lambda $, which are levitated in the presence of a strong magnetic
field (not to scale).}
\label{2-oil-drops-in-trap-4b}
\end{figure}

However, the ratio of the forces of gravity and electricity of two
\textquotedblleft Millikan oil drops\textquotedblright\ (to be described in
more detail below; see Figure \ref{2-oil-drops-in-trap-4b}) needs not be so
hopelessly small \cite{Lamb-medal}.

Suppose that each \textquotedblleft Millikan oil drop\textquotedblright\ has
a single electron attached firmly to it,\ and contains a Planck-mass amount
of superfluid helium, viz.,%
\begin{equation}
m_{\text{Planck}}=\sqrt{\frac{\hbar c}{G}}\approx 22\text{ micrograms}
\label{Planck-mass}
\end{equation}%
where $\hbar $ is Planck's constant/2$\pi $, $c$ is the speed of light, and $%
G$ is Newton's constant. Planck's mass sets the characteristic scale at
which quantum mechanics ($\hbar $) impacts relativistic gravity ($c$, $G$).
(Planck obtained this mass by means of dimensional analysis.) Note that the
extreme smallness of $\hbar $ compensates for the extreme largeness of $c$
and for the extreme smallness of $G$, so that this mass scale is \textit{%
mesoscopic}, and not astronomical, in size. This suggests that it may be
possible to perform some novel \textit{nonastronomical}, table-top-scale
experiments at the interface of quantum mechanics and general relativity,
which are accessible in the laboratory. Such experiments will be considered
here.

The forces of gravity and electricity between the two \textquotedblleft
Millikan oil drops\textquotedblright\ are exerted upon the centers of mass
and the centers of charge of the drops, respectively. Both of these centers
coincide with the geometrical centers of the spherical drops, assuming that
the charge of the electrons on the drops is uniformly distributed around the
outside surface of the drops in a spherically symmetric manner (like in an $%
S $ state). Therefore the ratio of the forces of gravity and electricity
between the two \textquotedblleft Millikan oil drops\textquotedblright\ now
becomes%
\begin{equation}
\frac{\left\vert F_{G}\right\vert }{\left\vert F_{e}\right\vert }=\frac{Gm_{%
\text{Planck}}^{2}}{e^{2}}=\frac{G\left( \hbar c/G\right) }{e^{2}}=\frac{%
\hbar c}{e^{2}}\approx 137\text{ .}  \label{137}
\end{equation}%
Now the force of gravity is approximately 137 times stronger than the force
of electricity, so that instead of a mutual repulsion between these two
charged, massive objects, there is now a mutual attraction between them. The
sign change from mutual repulsion to mutual attraction between these two
\textquotedblleft Millikan oil drops\textquotedblright\ occurs at a critical
mass $m_{\text{crit}}$ given by%
\begin{equation}
m_{\text{crit}}=\sqrt{\frac{e^{2}}{\hbar c}}m_{\text{Planck}}\approx 1.9%
\text{ micrograms}  \label{m_[crit]}
\end{equation}%
whereupon $\left\vert F_{G}\right\vert $ $=\left\vert F_{e}\right\vert $,
and the forces of gravity and electricity balance each other in equilibrium.
The radius of a drop with this critical mass of superfluid helium, which has
a density of $\rho =0.145$ g/cm$^{3}$, is%
\begin{equation*}
R=\left( \frac{3m_{\text{crit}}}{4\pi \rho }\right) ^{1/3}=146\text{
micrometers.}
\end{equation*}%
This is a strong hint that mesoscopic-scale quantum effects can lead to
non-negligible couplings between gravity and electromagnetism that can be
observed in the laboratory.

The critical mass $m_{\text{crit}}$ is also the mass at which there occurs a
comparable amount of generation of electromagnetic and gravitational
radiation power upon scattering of radiation from the pair of
\textquotedblleft Millikan oil drops,\textquotedblright\ each with a mass $%
m_{\text{crit}}$ and with a single electron attached to it. The ratio of
quadrupolar gravitational to the quadrupolar electromagnetic radiation power
is given by%
\begin{equation}
\frac{P_{GR}^{\text{(quad)}}}{P_{EM}^{\text{(quad)}}}=\frac{Gm_{\text{crit}%
}^{2}}{e^{2}}=1\text{ ,}  \label{Larmor-power-ratio}
\end{equation}%
where the factors of $\kappa $ in Equations (\ref{GR-Larmor}) and (\ref%
{EM-Larmor-quad}) cancel out, if the center of mass of each drop co-moves
rigidly together with its center of charge. This implies that the scattered
power from these two charged objects in the gravitational wave channel will
become equal to that in the electromagnetic wave channel. Note that a pair
of larger drops, whose masses have been increased beyond the critical mass,
will still satisfy Equation (\ref{Larmor-power-ratio}), provided that the
number of electrons on these drops is also increased proportionately so that
the charge-to-mass ratio of these drops remains fixed, and provided that the
system is placed in a strong magnetic field and cooled to low temperatures
so that it remains in the ground state.

\section{Maxwell-like equations that result from linearizing Einstein's
equations}

In order to understand the calculation of the scattering cross section of
the \textquotedblleft Millikan oil drops\textquotedblright\ to be given
below, let us start from a very useful Maxwell-like representation of the
linearized Einstein's equations of standard general relativity that
describes weak gravitational fields coupled to matter in the asymptotically
flat coordinate system of a distant inertial observer \cite{Wald}: 
\begin{equation}
\mathbf{\nabla \cdot E}_{G}=-\frac{\rho _{G}}{\varepsilon _{G}}
\label{Maxwell-like-eq-1}
\end{equation}%
\begin{equation}
\mathbf{\nabla \times E}_{G}=-\frac{\partial \mathbf{B}_{G}}{\partial t}
\label{Maxwell-like-eq-2}
\end{equation}%
\begin{equation}
\mathbf{\nabla \cdot B}_{G}=0  \label{Maxwell-like-eq-3}
\end{equation}%
\begin{equation}
\mathbf{\nabla \times B}_{G}=\mu _{G}\left( -\mathbf{J}_{G}+\varepsilon _{G}%
\frac{\partial \mathbf{E}_{G}}{\partial t}\right)  \label{Maxwell-like-eq-4}
\end{equation}%
where the gravitational analog of the magnetic permeability of free space is
given by%
\begin{equation}
\mu _{G}=\frac{4\pi G}{c^{2}}=9.31\times 10^{-27}\text{ SI units}
\label{mu_G}
\end{equation}%
and where the gravitational analog of the electric permittivity of free
space is given by%
\begin{equation}
\varepsilon _{G}=\frac{1}{4\pi G}=1.19\times 10^{9}\text{ SI units.}
\label{epsilon_G}
\end{equation}%
Taking the curl of the gravitational analog of Faraday's law, Equation (\ref%
{Maxwell-like-eq-2}), and substituting into its right side the gravitational
analog of Ampere's law, Equation (\ref{Maxwell-like-eq-4}), one obtains a
wave equation, which implies that the speed of gravitational radiation is
given by%
\begin{equation}
c=\frac{1}{\sqrt{\varepsilon _{G}\mu _{G}}}=3.00\times 10^{8}\text{ m/s,}
\label{speed-of-light}
\end{equation}%
which exactly equals the vacuum speed of light. In these Maxwell-like
equations, the field $\mathbf{E}_{G}$, which is the \emph{gravito-electric}
field, is to be identified with the local acceleration $\mathbf{g}$ of a
test particle produced by the mass density $\rho _{G}$, and the field $%
\mathbf{B}_{G}$, which is the \emph{gravito-magnetic} field produced by the
mass current density $\mathbf{J}_{G}$ and by the gravitational analog of the
Maxwell displacement current density $\varepsilon _{G}\partial \mathbf{E}%
_{G}/\partial t$, is to be identified with a time-dependent generalization
of the Lense-Thirring field of general relativity.

In addition to the speed $c$ of gravity waves, there is another important
physical property that these waves possess, which can be formed from the
gravito-magnetic permeability of free space $\mu _{G}$ and from the
gravito-electric permittivity $\varepsilon _{G}$ of free space, namely, the
characteristic impedance of free space $Z_{G}$, which is given by \cite%
{Kiefer-Weber}\cite{Chiao2004}%
\begin{equation}
Z_{G}=\sqrt{\frac{\mu _{G}}{\varepsilon _{G}}}=2.79\times 10^{-18}\text{ SI
units.}  \label{Z_G}
\end{equation}%
As in electromagnetism, this characteristic impedance of free space plays a
central role in all radiation problems, such as in a comparison of the
radiation resistance of gravity-wave antennas to the value of this impedance
in order to estimate the coupling efficiency of these antennas to free
space. The numerical value of this impedance is extremely small, but the
impedance of all material objects must be \textquotedblleft impedance
matched\textquotedblright\ to this extremely small quantity before
significant power can be transferred efficiently from gravity waves to these
objects, or vice versa.

However, all classical material objects, such as Weber bars, have such a
high dissipation and such a high radiation resistance that they are
extremely poorly impedance-matched to free space. They can therefore neither
absorb gravity wave energy, nor emit it efficiently \cite{Weinberg}\cite%
{Chiao2004}. Hence it is a common belief that all materials, whether
classical or quantum, are essentially completely transparent to
gravitational radiation.

Macroscopically coherent quantum matter, such as a quantum Hall fluid, can
be exceptions to this general rule, however, since they can be quantized so
as to have a strictly zero dissipation. In the case of the quantum Hall
effect, this \textquotedblleft quantum
dissipationlessness\textquotedblright\ arises from the large size of the
energy gap $E_{\text{gap}}=\hbar \omega _{\text{cycl}}$ where $\omega _{%
\text{cycl}}$ is the electron cyclotron frequency, when $E_{\text{gap}}$ is
compared with the small size of the thermal fluctuations due to $k_{B}T$ at
very low temperatures. The energy gap $E_{\text{gap}}$ is like the BCS gap
of superconductors \cite{Tinkham}. As in the case of superconductors, due to
the absence of excitations with energies within the energy gap, the
scattering of the electrons in the quantum Hall fluid by impurities and by
phonons in the material, is exponentially suppressed, and the quantum
many-body system thus becomes dissipationless. In the case of
superconductors, this is evidenced by the persistent electrical currents in
annular rings of superconductors that have been projected to last longer
than the age of the Universe.

Instead of discussing the case of superconductors here, however, we shall
focus instead on the case of quantum Hall fluids, since the proposed
experiments will not be utilizing superconductivity, but rather the quantum
Hall effect, for the coupling between electromagnetic and gravitational
radiations.

\section{Specular reflection of gravity waves by a quantum Hall fluid}

\bigskip A quantum Hall fluid consists of a two-dimensional electron gas
which forms at very low temperatures in the presence of a very strong
magnetic field. In solid-state physics, a quantum Hall fluid forms due to
the electrons trapped at the interface between two semiconductors, such as
gallium arsenide and gallium-aluminum arsenide, when the sample is cooled
down to milli-Kelvin-scale temperatures in the presence of Tesla-scale
magnetic fields. Experimental evidence that the quantum Hall fluid is
dissipationless comes from the fact that their quantum Hall plateaus are
extremely flat, in which, for example, the transverse Hall resistance is
quantized in exact integer multiples of $h/e^{2}$, where $h$ is Planck's
constant and $e$ is the electron charge, but the longitudinal Hall
resistance,\ which is responsible for dissipation, is quantized to become
exactly zero \cite{Prange}.

However, we shall be considering here the quantum Hall fluid that forms on
the surface of a superfluid helium drop. Impurity, phonon, roton, and
ripplon scattering of the electrons moving on the surface of the drop is
exponentially suppressed because of the essentially perfect superfluidity of
liquid helium at milli-Kelvin-scale temperatures. Thus the electrons can
slide frictionlessly along the surface of a \textquotedblleft Millikan oil
drop.\textquotedblright\ Since the electrons reside in a thin layer at a
very small distance of approximately 80 \AA\ away from the surface, which is
much smaller than the typical centimeter-scale size of the drops to be used
in the proposed experiments, locally the electronic motion is planar and can
be well approximated by the two-dimensional motion of an electron gas on a
frictionless dielectric plane (see Appendix A).

One important consequence of the zero-resistance property of a quantum Hall
fluid is that a mirror-like reflection of electromagnetic waves can occur at
a planar interface between the vacuum and the fluid. This reflection is
similar to that which occurs when an incident electromagnetic wave
propagates down a transmission line with a characteristic impedance $Z$,
which is then terminated by means of a resistor $R$ whose value is close to
zero. The reflection coefficient ${\mathcal{R}}$ of the wave from such a
termination is given by%
\begin{equation}
{\mathcal{R}}=\left\vert \frac{Z-R}{Z+R}\right\vert ^{2}\rightarrow 100\%%
\text{ when }R\rightarrow 0\text{ ,}
\label{Reflection-from-transmission-line}
\end{equation}%
which approaches arbitrarily close to 100\% when the resistance vanishes.
When the resistance $R=0$, low-frequency electromagnetic radiation fields
are \textquotedblleft shorted out\textquotedblright\ by the resistor $R$,
and specular reflection occurs.

From the Maxwell-like Equations (\ref{Maxwell-like-eq-1}) - (\ref%
{Maxwell-like-eq-4}), and the boundary conditions that follow from them \cite%
{Hossenfelder}, it follows that there should exist an analogous reflection
of a gravitational plane wave from a planar interface of the vacuum with the
quantum Hall fluid, whose reflection coefficient ${\mathcal{R}}_{G}$ is
given by%
\begin{equation}
{\mathcal{R}}_{G}=\left\vert \frac{Z_{G}-R_{G}}{Z_{G}+R_{G}}\right\vert
^{2}\rightarrow 100\%\text{ when }R_{G}\rightarrow 0\text{ .}
\label{Reflection-from-vacuum-superconductor-interface}
\end{equation}%
This counter-intuitive result arises from the fact that the quantum Hall
fluid can, under certain circumstances, possess a strictly zero dissipation,
and therefore an equivalent mass-current resistance $R_{G}$ that can also be
strictly zero, as compared to the characteristic impedance of free space $%
Z_{G}$ $=2.79\times 10^{-18}$ SI units given by Equation (\ref{Z_G}).
Although the gravitational impedance of free space $Z_{G\text{ }}$ is an
extremely small quantity, it is still a finite quantity. However, the
dissipative resistance of a quantum Hall fluid is quantized, and can
therefore be \textit{exactly} zero. When the resistance $R_{G}=0$,
low-frequency incident gravitational radiation fields are \textquotedblleft
shorted out\textquotedblright\ by $R_{G}$, and specular reflection occurs.

It may be objected that in Equation (\ref%
{Reflection-from-vacuum-superconductor-interface}), it is unclear exactly
how the thickness of the quantum Hall fluid compares in size relative to any
relevant \textquotedblleft penetration-depth\textquotedblright\ length
scales, and also that this Equation fails to take into account the
frequency-dependent complex impedance of the quantum Hall fluid. When taken
properly into account, it could have turned out that these effects would
have made the reflectivity ${\mathcal{R}}_{G}$ negligibly small. However,
when they are taken properly into account \cite{Tinkham2}, the result is
that although the reflectivity ${\mathcal{R}}_{G}$ is not strictly unity,
nevertheless it can be nonnegligible. The reflectivity ${\mathcal{R}}_{G}$
for gravity waves needs only be of the order of unity, and not strictly
unity, to be experimentally interesting.

Hence it follows that under certain circumstances to be spelled out below,
specular reflection of gravity waves can occur from a quantum Hall fluid.
Therefore mirrors for gravitational radiation in principle can exist. Curved
mirrors can focus this radiation, and Newtonian telescopes for gravity waves
can therefore in principle be constructed. In the case of scattering of
gravity waves from the \textquotedblleft Millikan oil
drops,\textquotedblright\ the above specular-reflection condition implies
hard-wall boundary conditions at the surfaces of these spheres, so that the
scattering cross section of these waves from a pair of large spheres can be
geometric, i.e., hard-sphere, in size.

However, one cannot tell whether these statements about specular reflection
of gravitational radiation from quantum Hall fluids are true or not
experimentally, without the existence of a source and a detector for such
radiation. The quantum transducers based on \textquotedblleft Millikan oil
drops\textquotedblright\ to be discussed in more detail below may provide
the needed source and detector.

Although we have been focusing in the above discussion on the case of the
quantum Hall fluid which forms on \textquotedblleft Millikan oil
drops,\textquotedblright\ we should remark that specular reflection of
gravity waves may also occur from a vacuum-superconductor interface. This
may possibly follow from the recent potentially very important discovery 
\cite{Tajmar} (which of course needs independent confirmation)\ that in an
angularly accelerating superconductor, such as a niobium ring rotating with
a steadily increasing angular velocity, there seems to be a large
enhancement of the gravito-magnetic field $\mathbf{B}_{G}$, apparently from
a macroscopically constructive quantum interference effect\ due to the
macroscopically coherent nature of the quantum mechanical phase of the
electrons in niobium, which in turn arises from the condensate of many
Cooper pairs of electrons in this superconductor. As a result of the angular
acceleration of the niobium ring, there seems to arise a steadily increasing
gravitational analog of the London moment in the form of a very large $%
\mathbf{B}_{G}$ field inside the ring, which is increasing linearly in time.
The gravitational analog of Faraday's law, Equation (\ref{Maxwell-like-eq-2}%
), then implies the generation of loops of the gravito-electric field $%
\mathbf{E}_{G}$ inside the hole of the ring, which can be detected by
sensitive accelerometers. The gravito-magnetic field $\mathbf{B}_{G}$ is
thus inferred to be many orders of magnitude greater than what one would
expect classically due to the mass current associated with the rigid
rotation of the ionic lattice of the ring. These observations have recently
been confirmed by replacing the electromechanical accelerometers with a
laser gyro \cite{Tajmar2}.

A tentative theoretical interpretation of these recent experiments is that
the coupling constant $\mu _{G}$ which couples the mass currents of the
superconductor to the gravito-magnetic field $\mathbf{B}_{G}$ is somehow
greatly enhanced due to the presence of the macroscopically coherent quantum
matter in niobium. This enhancement can be understood phenomenologically in
terms of a ferromagnetic-like enhancement factor $\kappa _{G}^{\text{(magn)}%
} $, which enhances the gravito-magnetic coupling constant \emph{inside the
medium} as follows:%
\begin{equation}
\mu _{G}^{\prime }=\kappa _{G}^{\text{(magn)}}\mu _{G}  \label{mu-G-prime}
\end{equation}%
where $\kappa _{G}^{\text{(magn)}}$ is a positive number much larger than
unity. This ferromagnetic-like enhancement factor $\kappa _{G}^{\text{(magn)}%
}$ is the gravitational analog of the magnetic permeability constant $\kappa
_{m}$ of ferromagnetic materials in the standard theory of electromagnetism.

The basic assumption of this phenomenological theory is that of a \emph{%
linear response} of the material medium to weak applied gravito-magnetic
fields \cite{Kramers-Kronig}; that is to say, whatever the fundamental
explanation is of the large observed positive values of $\kappa _{G}^{\text{%
(magn)}}$, the medium produces an enhanced gravito-magnetic field $\mathbf{B}%
_{G}$\ that is \emph{directly proportional} to the mass current density $%
\mathbf{J}_{G}$ of the ionic lattice. For weak fields, this is a reasonable
assumption. However, it should be noted that this phenomenological
explanation based on Equation (\ref{mu-G-prime}) is different from the
theoretical explanation based on Proca-like equations for gravitational
fields with a finite graviton rest mass, which was proposed by the
discoverers of the effect in Ref. \cite{Tajmar}.

Nevertheless, it is natural to consider introducing the phenomenological
Equation (\ref{mu-G-prime}) to explain the observations, since a large
enhancement factor $\kappa _{G}^{\text{(magn)}}$\ due to the material medium
is very similar to its analog in magnetism, which explains, for example, the
large ferromagnetic enhancement of the inductance of a solenoid by a
magnetically soft, permeable iron core with permeability $\kappa _{m}>>1$
that arises from the alignment of electron spins inside the iron. This
spin-alignment effect leads to the large observed values of the magnetic
susceptibility of iron, like those utilized in mu metal shields. Just as in
the case of the iron core inserted inside a solenoid, where the large
enhancement of the solenoid's inductance disappears above the Curie
temperature of iron, it was observed in these recent experiments that the
large gravito-magnetic enhancement effect disappears above the
superconducting transition temperature of niobium.

If the tentative phenomenological interpretation given by Equation (\ref%
{mu-G-prime}) of these experiments turns out to be correct, one important
consequence of the large resulting values of $\kappa _{G}^{\text{(magn)}}$
is that a mirror-like reflection should occur at a planar
vacuum-superconductor interface, where the refractive index of the
superconductor has an abrupt jump from unity to a value given by%
\begin{equation}
n_{G}=\left( \kappa _{G}^{\text{(magn)}}\right) ^{1/2}\text{ .}
\label{refractive-index}
\end{equation}

However, it should be immediately emphasized here that only positive masses
are observed to exist in nature, and not negative ones. Hence gravitational
analogs of electric dipole moments do not exist. It follows that the
gravitational analog $\kappa _{G}^{\text{(elec)}}$ of the usual dielectric
constant $\kappa _{e}$ for all kinds of matter, whether classical or
quantum, in the Earth's gravito-electric field $\mathbf{E}_{G}=\mathbf{g}$,
cannot differ from its vacuum value of unity, i.e.,%
\begin{equation}
\kappa _{G}^{\text{(elec)}}\equiv \varepsilon _{G}^{\prime }/\varepsilon
_{G}=1\text{ , }  \label{gravito-dielectric-constant=1}
\end{equation}%
exactly. Hence one cannot screen out, even partially, the gravito-electric
DC gravitational fields like the Earth's gravitational field using
superconducting Faraday cages. In particular, the local value of the
acceleration $\mathbf{g}$ due to Earth's gravity is not at all affected by
the presence of nearby matter with large $\kappa _{G}^{\text{(magn)}}$.

By contrast, the gravitational analog of Ampere's law combined with the
gravitational analog of the Lorentz force law \cite{Wald}%
\begin{equation}
\mathbf{F}_{G}=m\left( \mathbf{E}_{G}+4\mathbf{v\times B}_{G}\right) \text{ ,%
}  \label{Loretntz-Force-Law}
\end{equation}%
where $\mathbf{F}_{G}$ is the force on a test particle with mass $m$ and
velocity $\mathbf{v}$ (all quantities as seen by the distant inertial
observer), leads to the fact that a \textit{repulsive} component of force
exists between two parallel mass currents travelling in the same direction,
which is the opposite to the case in electricity, where two parallel
electrical currents travelling in the same direction \textit{attract} each
other \cite{Wald}. A repulsive \textit{gravito-magnetic} gravitational force
follows from the negative sign in front of the mass current density $\mathbf{%
J}_{G}$ in Equation (\ref{Maxwell-like-eq-4}), which is necessitated by the
conservation of mass, since upon taking the divergence of Equation (\ref%
{Maxwell-like-eq-4}), and combining it with Equation (\ref{Maxwell-like-eq-1}%
) (whose negative sign in front of the mass density $\rho _{G}$\ is fixed by
Newton's law of gravitation, where all masses \emph{attract} each other),
one must obtain the continuity equation for mass, i.e.,%
\begin{equation}
\mathbf{\nabla \cdot J}_{G}+\frac{\partial \rho _{G}}{\partial t}=0\text{ ,}
\label{continuity}
\end{equation}%
where $\mathbf{J}_{G}$ is the mass current density, and $\rho _{G}$\ is the
mass density. Moreover, the negative sign in front of the mass current
density $\mathbf{J}_{G}$ in the gravitational analog of Ampere's law,
Equation (\ref{Maxwell-like-eq-4}), implies an \textit{anti-Meissner}
effect, in which the lines of the $\mathbf{B}_{G}$\ field, instead of being
expelled from the superconductor, as in the usual Meissner effect, are
pulled tightly into the interior of the body of the superconductor when $%
\kappa _{G}^{\text{(magn)}}$\ is a large, positive number.

However, it should again be stressed that what is being proposed here in the
above phenomenological scenario does not at all imply an \textquotedblleft
anti-gravity\textquotedblright\ effect, in which the Earth's gravitational
field is somehow partially screened out by the so-called \textquotedblleft
Podkletnov effect\textquotedblright , where it was claimed that rotating
superconductors reduce by a few percent the gravito-electric field $\mathbf{E%
}_{G}=\mathbf{g}$, i.e., the local acceleration of all objects due to
Earth's gravity, in their vicinity. Experiments attempting to reproduce this
effect have failed to do so\ \cite{Tajmar}. The non-existence of the
\textquotedblleft Podkletnov effect\textquotedblright\ would be consistent
with the above phenomenological theory, since \textit{longitudinal}
gravito-electric fields cannot be screened under any circumstances; however, 
\textit{transverse} radiative gravitational fields can be reflected by
macroscopically coherent quantum matter.

Very large values of $\kappa _{G}^{\text{(magn)}}$ for superconductors would
imply that the index of refraction for gravitational plane waves in these
media would be considerably larger than unity, i.e.,%
\begin{equation}
n_{G}=\left( \kappa _{G}^{\text{(magn)}}\right) ^{1/2}\gtrsim 1\text{ .}
\label{index>>1}
\end{equation}%
The Fresnel reflection coefficient ${\mathcal{R}}_{G}$ of gravity waves
normally incident upon the vacuum-superconductor interface would therefore
become%
\begin{equation}
{\mathcal{R}}_{G}=\left\vert \frac{n_{G}-1}{n_{G}+1}\right\vert ^{2}\simeq 
\text{Order of unity,}  \label{Fresnel-reflection}
\end{equation}%
and could thus be large enough to be experimentally interesting. Again (but
for different reasons from those of Equation (\ref%
{Reflection-from-vacuum-superconductor-interface})), Equation (\ref%
{Fresnel-reflection}) would imply mirror-like reflection of these waves from
superconducting surfaces \cite{Tinkham2}. It should be noted that large
values of the ferromagnetic-like enhancement factor $\kappa _{G}^{\text{%
(magn)}}$, of the index of refraction $n_{G}$, and of the reflectivity ${%
\mathcal{R}}_{G}\,$, are not forbidden by the principle of equivalence,
which has been checked experimentally with extremely high accuracy, but only
within the gravito-electric sector of gravitation.

However, it should be emphasized here that although interesting and possibly
very important, the above discussion concerning superconductors as mirrors
for gravity waves is only secondary to the primary purpose of this paper,
which is to present the case for the possibility of efficient quantum
transducers via \textquotedblleft Millikan oil drops.\textquotedblright\
These kinds of quantum transducers do not involve the use of
superconductivity.

\section{\textquotedblleft Millikan oil drops\textquotedblright\ described
in more detail}

Let the oil of the classic Millikan oil drops be replaced with superfluid
helium ($^{4}$He) with a gravitational mass of around the
Planck-mass--scale, and let these drops be levitated in the presence of
strong, Tesla-scale magnetic fields.

The helium atom is diamagnetic, and liquid helium drops have successfully
been magnetically levitated in an anti-Helmholtz magnetic trapping
configuration \cite{Weilert1996}. Due to its surface tension, the surface of
a freely suspended, isolated, ultracold superfluid drop is ideally smooth,
i.e., atomically perfect, in the sense that there are no defects (such as
dislocations on the surface of an imperfect crystal) which can trap and
thereby localize the electron. The absence of any scattering centers for the
electrons on the surface of the superfluid helium of a \textquotedblleft
Millikan oil drop\textquotedblright\ implies that the electrons can move
frictionlessly, and hence dissipationlessly, over its surface.

When an electron approaches a drop, the formation of an image charge inside
the dielectric sphere of the drop causes the electron to be attracted by the
Coulomb force to its own image. As a result, it is experimentally observed
that the electron is bound\ to the outside surface of the drop in a
hydrogenic ground state. The binding energy of the electron to the surface
of liquid helium has been measured using millimeter-wave spectroscopy to be
8 Kelvin \cite{Grimes2}, which is quite large compared to the
milli-Kelvin-scale temperatures for the proposed experiments. Hence the
electron is tightly bound to the outside surface of the drop so that the
radial component of its motion is frozen, but when the drop becomes a
superfluid, the electron is free to move frictionlessly along its tangential
component of motion, and thus to become delocalized over the entire surface.

Such a \textquotedblleft Millikan oil drop\textquotedblright\ is a
macroscopically phase-coherent quantum object. In its ground state, which
possesses a single, coherent quantum mechanical phase throughout the
interior of the superfluid \cite{Footnote-A}, the drop possesses a zero
circulation quantum number (i.e., contains no quantum vortices), with one
unit (or an integer multiple) of the charge quantum number. As a result of
the drop being at ultra-low temperatures, all degrees of freedom other than
the center-of-mass degrees of freedom are frozen out, so that there results
a zero-phonon M\"{o}ssbauer-like effect, in which the entire mass of the
drop moves rigidly as a single unit in response to radiation fields (see
below). Therefore, the center of mass of the drop will co-move with the
center of charge. Also, since it remains adiabatically in the ground state
during perturbations due to these weak radiation fields, the
\textquotedblleft Millikan oil drop\textquotedblright\ possesses properties
of \textquotedblleft quantum rigidity\textquotedblright\ and
\textquotedblleft quantum dissipationlessness\textquotedblright\ that are
the two most important quantum properties for achieving a high coupling
efficiency for gravity-wave antennas \cite{Chiao2004}.

Note that two spatially separated \textquotedblleft Millikan oil
drops\textquotedblright\ with the same mass and charge\ have the correct
bilateral symmetry in order to couple to quadrupolar gravitational
radiation, as well as to quadrupolar electromagnetic radiation. The coupling
of the drops to dipolar electromagnetic radiation, however, vanishes due to
symmetry. When they are separated by a distance on the order of a
wavelength, they should become an efficient quadrupolar antenna capable of
generating, as well as detecting, gravitational radiation.

\begin{figure}[tbp]
\centerline{\includegraphics{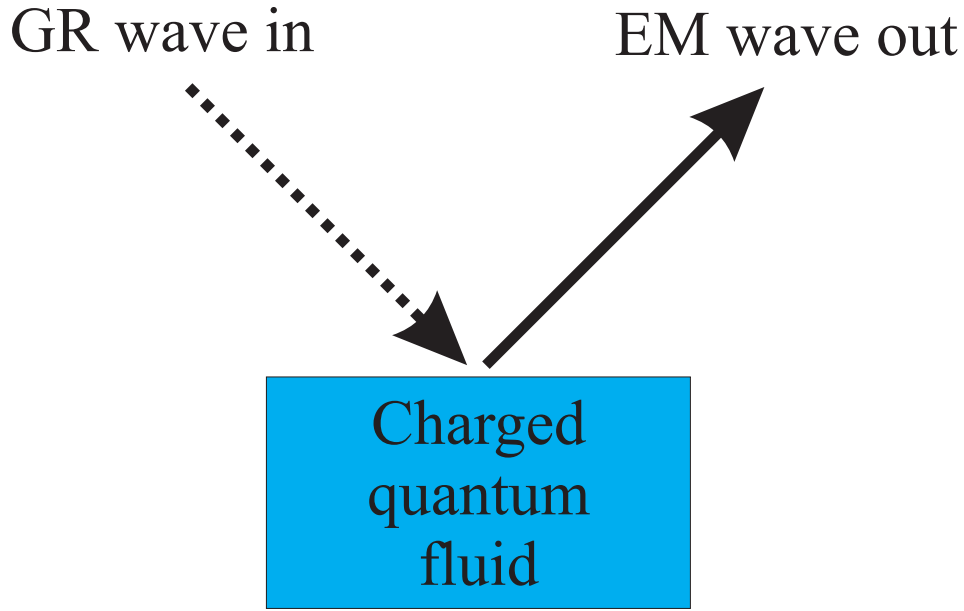}}
\caption{\textquotedblleft Charged quantum fluid\textquotedblright\ is a
quantum transducer consisting of a pair of \textquotedblleft Millikan oil
drops\textquotedblright\ in a strong magnetic field, which converts a
gravity (GR) wave into an electromagnetic (EM) wave.}
\label{GR-to-EM-wave-transducer-4b}
\end{figure}

\section{A pair of ``Millikan oil drops'' as a transducer}

Now imagine placing a pair of levitated \textquotedblleft Millikan oil
drops\textquotedblright\ separated by approximately a microwave wavelength
inside a black box, which represents a quantum transducer that can convert
gravitational (GR) waves into electromagnetic (EM) waves. See Figure \ref%
{GR-to-EM-wave-transducer-4b}. This kind of transducer action is similar to
that of the tidal force of a gravity wave passing over a pair of charged,
freely falling objects orbiting the Earth, which can in principle convert a
GR wave into an EM\ wave \cite{Lamb-medal}. Such transducers are linear,
reciprocal devices.

By time-reversal symmetry \cite{Yu}, the reciprocal process, in which
another pair of \textquotedblleft Millikan oil drops\textquotedblright\
converts an EM wave back into a GR wave, must occur with the same efficiency
as the forward process, in which a GR wave is converted into an EM wave by
the first pair of \textquotedblleft Millikan oil drops.\textquotedblright\
The time-reversed process is important because it allows the \emph{generation%
} of gravitational radiation, and can therefore become a practical source of
such radiation. The radiation reaction or back-action by the EM fields upon
the GR fields via these coherent quantum mechanical drops leads necessarily
to a non-negligible reciprocal process of the generation of these fields.
These actions must be mutual ones between these two kinds of radiation
fields.

\begin{figure}[tbp]
\centerline{\includegraphics{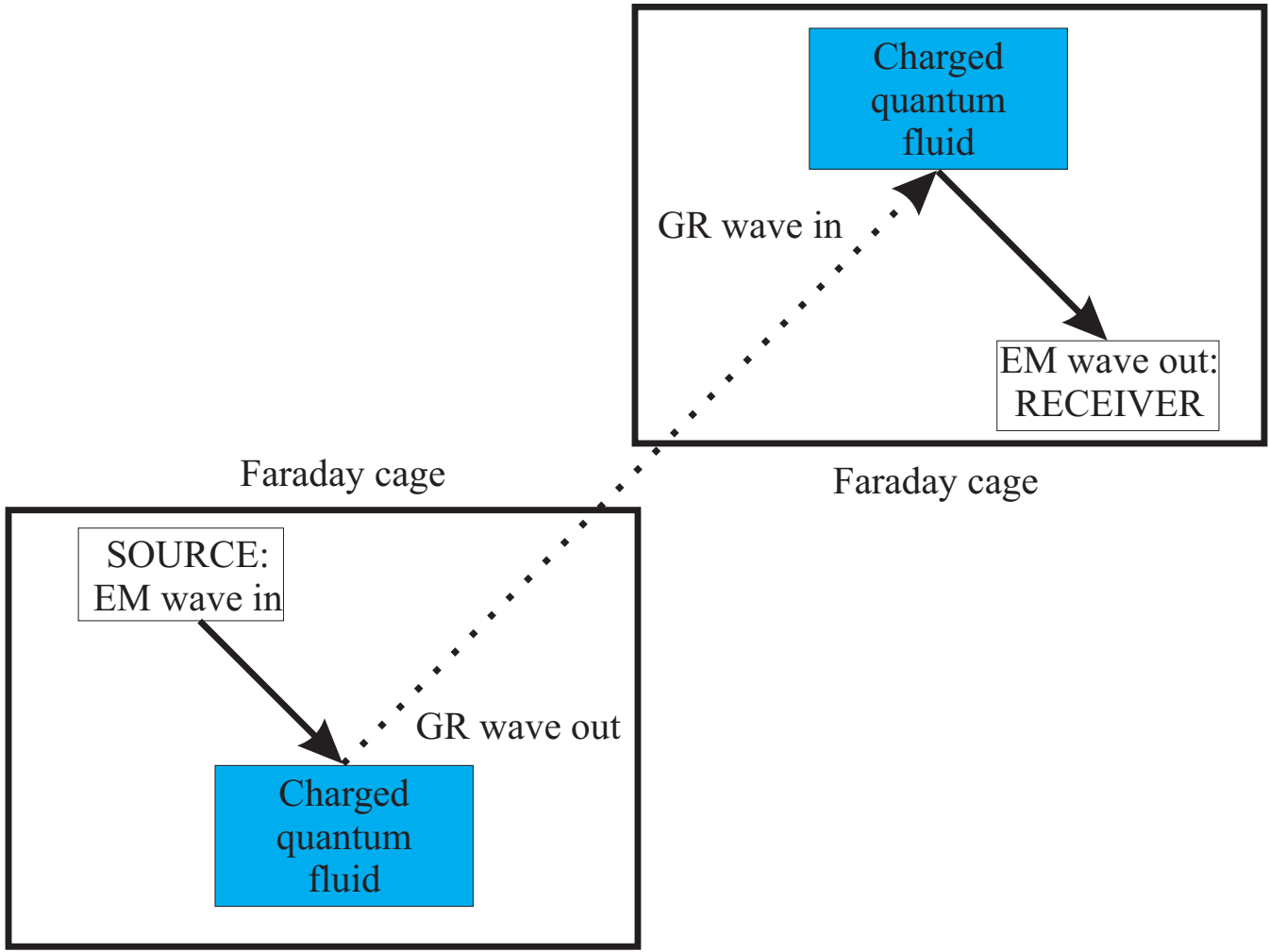}}
\caption{A Hertz-like experiment, in which EM waves are converted by the
lower-left quantum transducer (\textquotedblleft Charged quantum
fluid\textquotedblright ) into GR waves at the source, and the GR waves thus
generated are back-converted back into EM waves by the upper-right quantum
transducer at the receiver. Communication by EM waves is prevented by the
normal (i.e., nonsuperconducting) Faraday cages.}
\label{Hertz-4b}
\end{figure}

This raises the possibility of performing a Hertz-like experiment, in which
the time-reversed quantum transducer process becomes the source, and its
reciprocal quantum transducer process becomes the receiver of GR waves. See
Figure \ref{Hertz-4b}. Faraday cages consisting of nonsuperconducting metals
prevent the transmission of EM waves, so that only GR waves, which can
easily pass through all classical matter such as the normal (i.e.,
dissipative)\ metals of which standard, room-temperature Faraday cages are
composed, are transmitted between the two halves of the apparatus that serve
as the source and the receiver, respectively. Such an experiment would be
practical to perform using standard microwave sources and receivers, since
the scattering cross-sections and the transducer conversion efficiencies of
the two \textquotedblleft Millikan oil drops\textquotedblright\ turn out not
to be too small, as will be shown below.

\section{M{\"{o}}ssbauer-like response of \textquotedblleft Millikan oil
drops\textquotedblright\ in strong magnetic fields to radiation fields}

Let a pair of levitated \textquotedblleft Millikan oil
drops\textquotedblright\ be placed in strong, Tesla-scale magnetic fields,
and let the drops be separated by a distance on the order of a microwave
wavelength, which is chosen so as to satisfy the impedance-matching
condition for a good quadrupolar microwave antenna.

Now let a beam of electromagnetic waves in the Hermite-Gaussian TEM$_{11}$
mode \cite{Yariv1967}, which has a quadrupolar transverse field pattern that
has a substantial overlap with that of a gravitational plane wave, impinge
at a 45$^{\circ }$ angle with respect to the line joining these two charged
objects. Such a mode has been successfully generated using a
\textquotedblleft T\textquotedblright -shape microwave antenna \cite%
{Chiao2004}. As a result of being thus irradiated, the pair of
\textquotedblleft Millikan oil drops\textquotedblright\ will be driven into
relative motion in an anti-phased manner, so that the distance between them
will oscillate sinusoidally with time, according to an observer at infinity.
Thus the simple harmonic motion of the two drops relative to one another (as
seen by this observer) produces a time-varying mass quadrupole moment at the
same frequency as that of the driving electromagnetic wave. This oscillatory
motion will in turn scatter (in a linear scattering process) the incident
electromagnetic wave into gravitational and electromagnetic scattering
channels with comparable powers, provided that the ratio of quadrupolar
radiation powers is that given by Equation (\ref{Larmor-power-ratio}), i.e.,
is of the order of unity, which will be case if the charge-to-mass ratio of
the drops is the same as that of a single electron on a drop with a critical
mass $m_{\text{crit}}$. The reciprocal scattering process will also have a
power ratio of the order of unity. Pairs of large superfluid drops with many
electrons on them can be used as scatterers, as long as their charge-to-mass
ratio is consistent with Equation (\ref{Larmor-power-ratio}).

The M{\"{o}}ssbauer-like response of \textquotedblleft Millikan oil
drops\textquotedblright\ will now be discussed in more detail.\ Imagine what
would happen if one were to replace an electron in the vacuum with a single
electron which is firmly attached to the outside surface of a drop of
superfluid helium in the presence of a strong magnetic field and at ultralow
temperatures, so that the system of the electron and the superfluid,
considered as a single quantum entity, would form a single, macroscopic
quantum ground state \cite{Gigantic-atom}. Such a quantum system can possess
a sizeable gravitational mass. For the case of many electrons attached to a
large, massive drop, where a quantum Hall fluid forms on the outside surface
of the drop in the presence of a strong magnetic field, there results a
Laughlin-like ground state, which is the many-body state of an
incompressible quantum fluid \cite{Laughlin}. The property of quantum
incompressibility of such a fluid is equivalent to the property of
\textquotedblleft quantum rigidity,\textquotedblright\ which is one
necessary requirement for achieving high efficiency in
gravitational-radiation antennas, as was pointed out in \cite{Chiao2004}.
Like superfluids and superconductors, this fluid is also frictionless, i.e.,
dissipationless. This fulfills the condition of \textquotedblleft quantum
dissipationlessness,\textquotedblright\ which is another necessary
requirement for the successful construction of efficient gravity-wave
antennas \cite{Chiao2004}.

In the presence of strong, Tesla-scale magnetic fields, an electron is
prevented from moving at right angles to the local magnetic field line
around which it is executing tight cyclotron orbits. The result is that the
surface of the drop, to which the electron is tightly bound, cannot undergo
low-frequency liquid-drop deformations, such as the oscillations between the
prolate and oblate spheroidal configurations of the drop which would occur
at low frequencies in the absence of the magnetic field. After the drop has
been placed into Tesla-scale magnetic fields at milli-Kelvin-scale operating
temperatures, both the single- and many-electron drop systems will be
effectively frozen into the ground state, since the characteristic energy
scale for electron cyclotron motion in Tesla-scale fields is on the order of
Kelvins. Due to the tight coupling of the electron(s) to the outside surface
of the drop, also on the scale of Kelvins, this would effectively freeze out
all low-frequency shape deformations of the superfluid drop.

Since all internal degrees of freedom of the drop, such as its microwave
phonon excitations, will also be frozen out at sufficiently low
temperatures, the charge and the entire mass of the \textquotedblleft
Millikan oil drop\textquotedblright\ will co-move rigidly together as a
single unit, in a zero-phonon, M\"{o}ssbauer-like response to applied
radiation fields with frequencies below the cyclotron frequency. This is a
result of the elimination of all internal degrees of freedom by the
Boltzmann factor at sufficiently low temperatures, so that the system stays
in its ground state, and only the external degrees of freedom of the drop,
consisting only of its center-of-mass motions, remain.

The criterion for this zero-phonon, or M\"{o}ssbauer-like, mode of response
of the electron-drop system is that the temperature of the system is
sufficiently low, so that the probability for the entire system to remain in
its ground state without even a single quantum of excitation of any of its
internal degrees of freedom being excited, is very high, i.e.,%
\begin{equation}
\text{Prob. of zero internal excitation}\approx 1-\exp \left( -\frac{E_{%
\text{gap}}}{k_{B}T}\right) \rightarrow 1\text{ as }\frac{k_{B}T}{E_{\text{%
gap}}}\rightarrow 0,  \label{Prob(no excitation)}
\end{equation}%
where $E_{\text{gap}}$ is the energy gap separating the ground state from
the lowest permissible excited states, $k_{B}$ is Boltzmann's constant, and $%
T$ is the temperature of the system. Then the quantum adiabatic theorem
ensures that the system will stay adiabatically in the ground state of this
quantum many-body system during adiabatic perturbations, such as those due
to weak, externally applied radiation fields with frequencies below the
cyclotron frequency. By momentum conservation, since there are no internal
excitations to take up the radiative momentum transfer, the center of mass
of the entire system must undergo recoil in the emission and absorption of
radiation. Thus the mass involved in the response to radiation fields is the
entire mass of the whole system.

For the case of a single electron (or many electrons in the case of the
quantum Hall fluid)\ in a strong magnetic field, the typical energy gap is
given by%
\begin{equation}
E_{\text{gap}}=\hbar \omega _{\text{cycl}}=\frac{\hbar eB}{m}>>k_{B}T\text{ ,%
}  \label{Cyclotron-gap}
\end{equation}%
where $\omega _{\text{cycl}}=eB/m$ is the electron cyclotron frequency in SI
units. This inequality is valid for the Tesla-scale fields and
milli-Kelvin-scale temperatures in the experiments being considered here.

\section{Estimate of the scattering cross-section}

Let $d\sigma _{a\rightarrow \beta }$ be the differential cross-section for
the scattering of a mode $a$ of radiation of an incident gravitational wave
to a mode $\beta $ of a scattered electromagnetic wave by a pair of
\textquotedblleft Millikan oil drops\textquotedblright\ (Latin subscripts
denote GR waves, and Greek subscripts EM waves). Then, by time-reversal
symmetry \cite{Yu}%
\begin{equation}
d\sigma _{a\rightarrow \beta }=d\sigma _{\beta \rightarrow a}\text{ .}
\end{equation}%
Since electromagnetic and weak gravitational fields both formally obey
Maxwell's equations (apart from a difference in the signs of the source
density and the source current density; see Equations (\ref%
{Maxwell-like-eq-1}) - (\ref{Maxwell-like-eq-4})), and since these fields
obey the same boundary conditions \cite{Hossenfelder}\cite{Tinkham2}, the
solutions for the modes for the two kinds of scattered radiation fields must
also have the same mathematical form. Let $a$ and $\alpha $ be a pair of
corresponding solutions, and $b$ and $\beta $ be a different pair of
corresponding solutions to Maxwell's equations for GR and EM modes,
respectively. For example, $a$ and $\alpha $ could represent incoming plane
waves which copropagate in the same direction, and $b$ and $\beta $
scattered, outgoing plane waves which copropagate together in a different
direction. Then for a pair of drops with the same charge-to-mass ratio as
that for critical-mass drops with single electrons, there is an equal
conversion into the two types of scattered radiation fields in accordance
with Equation (\ref{Larmor-power-ratio}), and therefore%
\begin{equation}
d\sigma _{a\rightarrow b}=d\sigma _{a\rightarrow \beta }\text{ ,}
\end{equation}%
where $b$ and $\beta $ are corresponding modes of the two kinds of scattered
radiations.

By the same line of reasoning, for this pair of drops%
\begin{equation}
d\sigma _{b\rightarrow a}=d\sigma _{\beta \rightarrow a}=d\sigma _{\beta
\rightarrow \alpha }\text{ .}
\end{equation}%
It therefore follows from the principle of reciprocity (i.e., detailed
balance or time-reversal symmetry) that%
\begin{equation}
d\sigma _{a\rightarrow b}=d\sigma _{\alpha \rightarrow \beta }.
\end{equation}

In order to estimate the size of the total cross-section, it is easier to
consider first the case of electromagnetic scattering, such as the
scattering of microwaves from a pair of large drops with radii $R$ and a
separation $r$ on the order of a microwave wavelength (but with $r>2R$). The
diameter $2R$ of the drops can be made to be comparable to their separation $%
r\simeq \lambda $, (e.g., with $2\pi R=\lambda $ for the first Mie
resonance), provided that many electrons are added on their surfaces, so
that their charge-to-mass ratio is maintained to be the same as that of a
single electron on a critical-mass drop (this requires the addition of 20
thousand electrons for the first Mie resonance at $\lambda =2.5$ cm, where $%
R=4$ mm), and therefore Equation (\ref{Larmor-power-ratio}) still holds for
these large drops.

For an incident EM wave of a particular circular polarization, even just a
single, delocalized electron in the presence of a strong magnetic field is
enough to produce specular reflection of this wave (see Appendix A).
Therefore for circularly polarized light, the two drops behave like
perfectly conducting, shiny, mirrorlike spheres, which scatter light in a
manner similar to that of perfectly elastic hard-sphere scattering in
idealized billiards. The total cross section for the scattering of
electromagnetic radiation from this pair of large drops is therefore given
approximately by the geometric cross-sectional areas of two hard spheres%
\begin{equation}
\sigma _{\alpha \rightarrow \text{all }\beta }=\int d\sigma _{\alpha
\rightarrow \beta }\simeq \text{Order of }\pi R^{2}
\label{Geometric-X-section}
\end{equation}%
where $R$ is the hard-sphere radius of a drop. This hard-sphere
cross-section is much larger than the Thomson cross-section for the
classical, \emph{localized} single free-electron scattering of
electromagnetic radiation.

However, if, as one might expect on the basis of the prevailing (but
possibly incorrect) opinion that all gravitational interactions with matter,
including the scattering of gravitational waves from all types of matter, is
completely independent of whether this matter is classical or
quantum-mechanical in nature on any scale of size, and that therefore the
scattering cross-section for the drops would be extremely small as it is for
the classical Weber bar, then by reciprocity, the total cross-section for
the scattering of electromagnetic waves from the two-drop system must also
be extremely small. In other words, if \textquotedblleft Millikan oil
drops\textquotedblright\ were to be essentially invisible to gravitational
radiation as is commonly believed, then by reciprocity they must also be
essentially invisible to electromagnetic radiation. To the contrary, if it
should turn out that the quantum Hall fluid on the surface of these drops
should make them behave like superconducting spheres, then the earlier
discussion in connection with Equation (\ref%
{Reflection-from-vacuum-superconductor-interface}) would imply that the
total cross-section of these drops will be like that of hard-sphere
scattering, so that they certainly would not be invisible.

\section{A proposed preliminary experiment}

\begin{figure}[tbp]
\centerline{\includegraphics{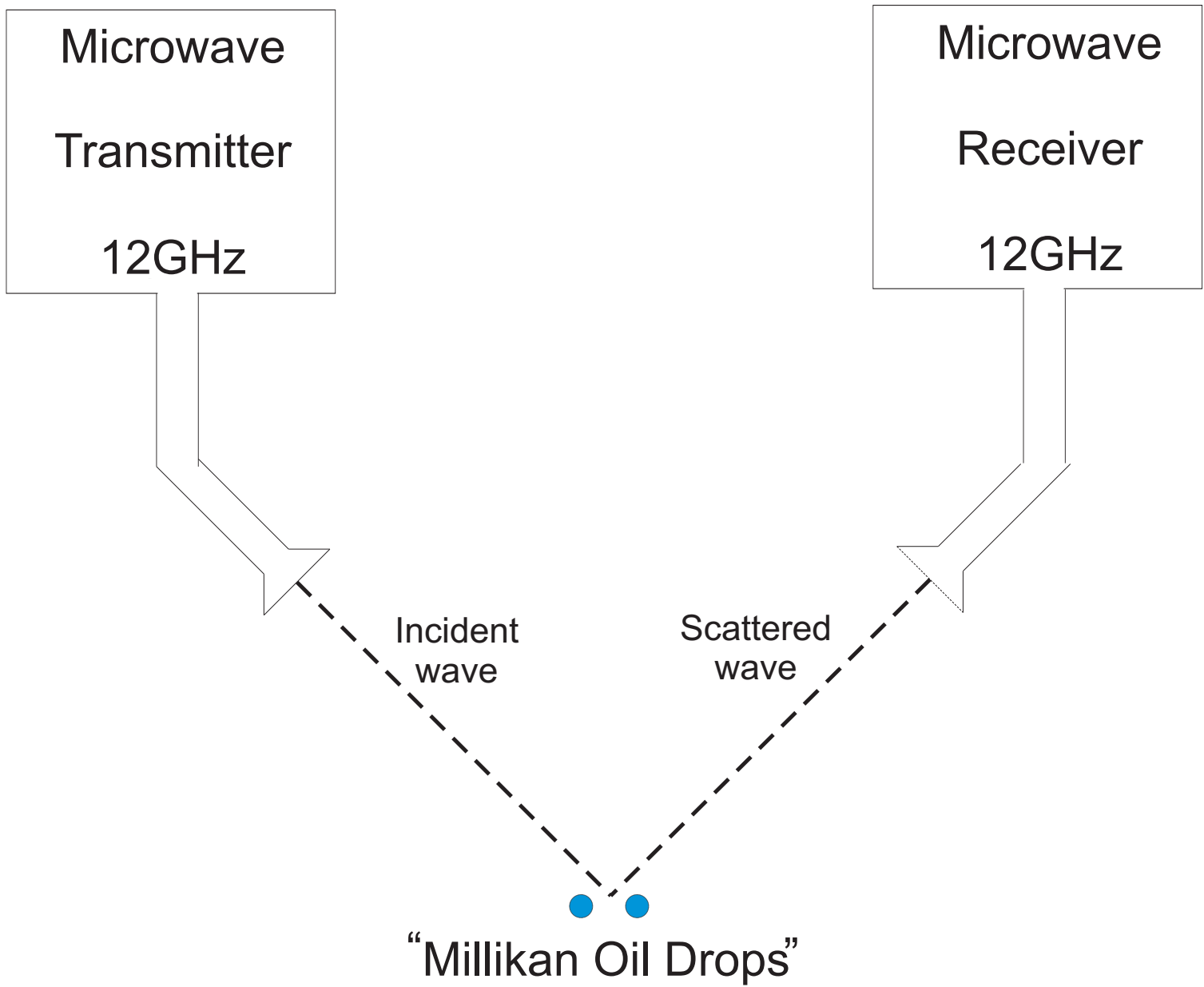}}
\caption{Schematic of apparatus (not to scale) to measure the scattering
cross-section of quadrupolar microwaves from a pair of \textquotedblleft
Millikan oil drops\textquotedblright\ in a strong magnetic field at low
temperatures.}
\label{Rescaled-scattering-experiment-4}
\end{figure}

In order to check the above hard-sphere scattering cross-section result, we
propose to first perform in a preliminary experiment a measurement of the
scattering cross section for quadrupolar microwave radiation off of a pair
of large \textquotedblleft Millikan oil drops\textquotedblright\ (see Figure %
\ref{Rescaled-scattering-experiment-4}). A standard oscillator at 12 GHz
emits microwaves which are prepared in a quadrupolar TEM$_{11}$ mode and
directed in a beam towards these drops, which are placed in a large magnetic
field and cooled to ultralow temperatures. \ The intensity of the scattered
microwave beam generated by the pair of drops is then measured by means of a
standard 12 GHz heterodyne receiver, which receives a quadrupolar TEM$_{11}$
mode. The purpose of this experiment is the check if the scattering
cross-section is indeed as large as the geometric cross-section predicted by
Equations (\ref{Reflection-from-vacuum-superconductor-interface}), (\ref%
{Geometric-X-section}), and (\ref{Fresnel-Reflection-for-plasma}). As one
increases the temperature, one should observe the disappearance of this
enhanced scattering cross section above the quantum Hall transition
temperature or the superfluid lambda point, whichever comes first.

\section{A common misconception corrected}

In connection with the idea that an EM wave incident on a pair of drops
could generate a GR wave, there arises a common misconception that the drops
are so heavy that their large inertia will prevent them from moving with any
appreciable amplitude in response to the driving EM wave amplitude. How can
they then possibly generate copious amounts of GR waves? This objection
overlooks the major role played by the principle of equivalence in the
motion of the drops, as will be explained below.

According to the equivalence principle, two tiny inertial observers, who are
undergoing free fall, i.e., who are freely floating near their respective
centers of the two \textquotedblleft Millikan oil drops,\textquotedblright\
would see no acceleration at all of the nearby surrounding matter of their
drop (nor would they feel any forces) due to the gravitational fields
arising from a gravity wave passing over the two drops. However, when they
measure the distance separating the two drops, by means of laser
interferometry, for example, they would conclude that the other drop is
undergoing acceleration relative to their drop, due to the fact that the 
\textit{space} between the drops is being alternately stretched and squeezed
by the incident gravity wave.\ They would therefore further conclude that
the charges attached to the surfaces of their locally freely-falling drops
would radiate electromagnetic radiation, in agreement with the observations
of the observer at infinity, who sees two charges undergoing time-varying
relative acceleration in response to the passage of the gravity wave.

According to the reciprocity principle, this scattering process can be
reversed in time. Under time reversal, the scattered electromagnetic wave
now becomes a wave which is incident on the drops. Again, the two tiny
inertial observers near the center of the drops would see no acceleration at
all of the surrounding matter (nor would they feel any forces) due to the
electric and magnetic fields of the incident electromagnetic wave. Rather,
they would conclude from measurements of the distance separating the two
drops, that it is again the \textit{space} between the drops that is being
alternately squeezed and stretched by the incident electromagnetic wave.
They would again further conclude that the masses associated with their
locally freely-falling drops would radiate gravitational radiation, in
agreement with the observations of the observer at infinity, who sees two
masses undergoing time-varying relative acceleration in response to the
passage of the electromagnetic wave.

From this general relativistic viewpoint, which is based upon the
equivalence principle, the fact that the drops might possess very large
inertias is irrelevant, since in fact the drops are not moving at all with
respect to the local inertial observer located at the center of drop.
Instead of causing motion of the drops \emph{through} space, the
gravitational fields of the incident gravitational wave are acting directly 
\emph{upon} space itself by alternately stretching and squeezing the space
in between the drops. Likewise, in the reciprocal process the very large
inertias of the drops are again irrelevant, since the electromagnetic wave
is not producing any motion at all of these drops with respect to the same
inertial observer \cite{Footnote-B}. Instead of causing motion of the drops 
\emph{through} space, the electric and magnetic fields of the incident
electromagnetic wave are again acting directly \emph{upon} space itself by
alternately squeezing and stretching the space in between the drops. The
time-varying, accelerated motion of the drops as seen by the distant
observer that causes quadrupolar radiation to be emitted in both cases, is
due to the time-varying \textit{curvature} of spacetime induced both by the
incident gravitational wave and by the incident electromagnetic wave. It
should be remembered that the space inside which the drops reside is
therefore no longer flat, so that the Newtonian concept of a
radiation-driven, local accelerated motion of a heavy drop through a fixed,
flat Euclidean space, is therefore no longer valid.

\section{The strain of space produced by the drops for a milliwatt of GR
wave power}

Another common objection to these ideas is that the strain of space produced
by a milliwatt of an electromagnetic wave is much too small to detect.
However, in the Hertz-like experiment, one is not trying to detect directly
the \textit{strain} of space (as in LIGO), but rather the \textit{power}
that is being transferred by the gravitational radiation fields from the
source to the receiver.

Let us put in some numbers. Suppose that one succeeded in completely
converting a milliwatt of EM wave power into a milliwatt of GR wave power at
the source. How big a strain amplitude of space would be produced by the
resulting GR wave? The gravitational analog of the time-averaged Poynting
vector is given by \cite{Weinberg}%
\begin{equation}
\left\langle S\right\rangle =c\left\langle t_{\mu \nu }\right\rangle =\frac{%
\omega ^{2}c^{3}}{8\pi G}h_{+}^{2}
\end{equation}%
where $\left\langle t_{\mu \nu }\right\rangle $ are certain components of
the time-averaged stress-energy tensor of a plane wave and $h_{+}$ is the
dimensionless strain amplitude of space for one polarization of a
monochromatic plane wave. For a milliwatt of power in such a plane wave at
30 GHz focused by means of a Newtonian telescope to a 1 cm$^{2}$ Gaussian
beam waist, one obtains a dimensionless strain amplitude of%
\begin{equation}
h_{+}\simeq 2\times 10^{-24}.
\end{equation}%
This strain is indeed exceedingly difficult to directly detect. \ However,
it is not necessary to directly measure the strain of space in order to
detect gravitational radiation, just as it is not necessary to directly
measure the electric field of a light wave, which may also be exceedingly
small, in order to be able to detect this wave. \ Instead, one can measure
directly the power conveyed by a beam of light by means of bolometry, for
example. \ Likewise, if one were to succeed to completely back-convert this
milliwatt of GR wave power back into a milliwatt of EM power at the
receiver, this amount of power would be easily detectable by standard
microwave techniques.

\section{Signal-to-noise considerations}

The signal-to-noise ratio expected for the Hertz-like experiment depends on
the current status of microwave source and receiver technologies. Based on
the experience gained from the experiment done on YBCO using existing
off-the-shelf microwave components \cite{Chiao2004}, we expect that we would
need geometric-sized cross-sections and a minimum conversion efficiency on
the order of a few parts per million per transducer, in order to detect a
signal. The overall system's signal-to-noise ratio depends on the initial
microwave power, the scattering cross-section, the conversion efficiency of
the quantum transducers, and the noise temperature of the microwave receiver
(i.e., its first-stage\ amplifier).

Microwave low-noise amplifiers can possess noise temperatures that are
comparable to room temperature (or even better, such as in the case of
liquid-helium cooled paramps or masers used in radio astronomy). The minimum
power $P_{\min }$ detectable in an integration time $\tau $ is given by%
\begin{equation}
P_{\min }=\frac{k_{B}T_{\text{noise}}\Delta \nu }{\sqrt{\tau \Delta \nu }}
\end{equation}%
where $k_{B}$ is Boltzmann's constant, $T_{\text{noise}}$ is the noise
temperature of the first stage microwave amplifier, and $\Delta \nu $ is its
bandwidth. Assuming an integration time of one second, and a bandwidth of 1
GHz, and a noise temperature $T_{\text{noise}}=300$ K, one gets $P_{\min
}(\tau =$1 sec$)=1.3\times 10^{-25}$ Watts, which is much less than the
milliwatt power levels of typical microwave sources.

\section{Possible applications}

\bigskip If we should be successful in the Hertz-like experiment, this could
lead to important possible applications in science and engineering. In
science, it would open up the possibility of gravity-wave astronomy at
microwave frequencies. One important problem to explore would be
observations of the analog of the Cosmic Microwave Background (CMB) in
gravitational radiation. Since the Universe is much more transparent to
gravity waves than to electromagnetic waves, such observations would allow a
much more penetrating look into the extremely early Big Bang towards the
Planck scale of time, than the presently well-studied CMB. \ Different
cosmological models of the very early Universe give widely differing
predictions of the spectrum of this penetrating radiation, so that by
measurements of the spectrum, one could tell which model, if any, is close
to the truth \cite{NASA2006}. The anisotropy in this radiation would also be
very important to observe.

In engineering, it would open up the possibility of intercontinental
communication by means of microwave-frequency gravity waves directly through
the interior of the Earth, which is transparent to such waves. This would
eliminate the need of communications satellites, and would allow
communication with people deep underground or underwater in submarines in
the Oceans. Such a new direction of gravity-wave engineering could aptly be
called \textquotedblleft gravity radio\textquotedblright .

\section{Appendix A: Specular reflection of a circularly polarized EM wave
by a delocalized electron moving on a plane in the presence of a strong
magnetic field}

Here we address the question: What is the critical frequency for specular
reflection of an EM plane wave normally incident upon a plane, in which
electrons are moving in the presence of a strong B field?\ \ The motivation
for solving this problem is to answer also the following questions: How can
just a single electron on the outside surface of a \textquotedblleft
Millikan oil drop\textquotedblright\ generate enough current in response to
an incident EM wave, so as to produce a re-radiated wave which totally
cancels out the incident wave within the interior of the drop, with the
result that none of the incident radiation can enter into the drop? Why does
specular reflection occur from the surface of such a drop, and hence why
does a hard-sphere EM cross-section result for a pair of \textquotedblleft
Millikan oil drops\textquotedblright ?

To simplify this problem to its bare essentials, let us examine first a
simpler, planar problem consisting of a uniform electron gas moving
classically on a frictionless, planar dielectric surface. We shall start
from a 3D point of view, but the Coulombic attraction of the electrons to
their image charges inside the dielectric will confine them in the direction
normal to the plane, so that the electrons are restricted to a 2D motion,
i.e., to motion in the two transverse dimensions of the plane. The electrons
are subjected to a strong DC magnetic field applied normally to this plane.
What is the linear response of this electron gas to a weak, normally
incident EM plane wave? Does a specular plasma-like reflection occur below a
critical frequency, even when just only a single, delocalized electron is
present on the plane? Let us first solve this problem classically.

Let the plane in question be the $z=0$ plane, and let a strong, applied DC $%
\mathbf{B}$ field be directed along the positive $z$ axis. The Lorentz force
on an electron is given by%
\begin{equation}
\mathbf{F}=e\left( \mathbf{E}+\frac{\mathbf{v}}{c}\mathbf{\times B}\right)
\end{equation}%
where $\mathbf{E}$, the weak electric field of the normally incident plane
wave, lies in the $(x,y)$ plane. (We shall use Gaussian units here.) The
cross product $\mathbf{v\times B}$ is given by%
\begin{equation}
\mathbf{v\times B=}\left\vert 
\begin{array}{ccc}
\mathbf{i} & \mathbf{j} & \mathbf{k} \\ 
v_{x} & v_{y} & 0 \\ 
0 & 0 & B%
\end{array}%
\right\vert =\mathbf{i}v_{y}B-\mathbf{j}v_{x}B~.
\end{equation}%
Hence Newton's equations of motion reduce to $x$ and $y$ components only%
\begin{equation}
F_{x}=m\ddot{x}=eE_{x}+\frac{v_{y}}{c}eB=eE_{x}+\frac{\dot{y}}{c}eB
\label{Newton-x}
\end{equation}%
\begin{equation}
F_{y}=m\ddot{y}=eE_{y}-\frac{v_{x}}{c}eB=eE_{x}-\frac{\dot{x}}{c}eB\text{ .}
\label{Newton-y}
\end{equation}%
Let us assume that the driving plane wave is a weak monochromatic wave with
the exponential time dependence%
\begin{equation}
E=E_{0}\exp \left( -i\omega t\right) ~.
\end{equation}%
Then assuming a linear response of the system to the weak incident EM wave,
the displacement, velocity, and acceleration of the electron all have the
same exponential time dependence%
\begin{equation}
x=x_{0}\exp \left( -i\omega t\right) \text{ and }y=y_{0}\exp \left( -i\omega
t\right)
\end{equation}%
\begin{equation}
\dot{x}=\left( -i\omega \right) x\text{ and }\dot{y}=\left( -i\omega \right)
y
\end{equation}%
\begin{equation}
\ddot{x}=-\omega ^{2}x\text{ and }\ddot{y}=-\omega ^{2}y
\end{equation}%
which converts the two ODEs, Equations (\ref{Newton-x}) and (\ref{Newton-y}%
), into the two algebraic equations for \ $x$ and $y$%
\begin{equation}
-m\omega ^{2}x=eE_{x}-\frac{i\omega y}{c}eB
\end{equation}%
\begin{equation}
-m\omega ^{2}y=eE_{y}+\frac{i\omega x}{c}eB\text{ .}
\end{equation}%
Let us now add $\pm i$ times the second equation to the first equation.
Solving for $x\pm iy$, one gets%
\begin{equation*}
x\pm iy=e\left( \frac{E_{x}\pm iE_{y}}{-m\omega ^{2}\pm \omega eB/c}\right)
\end{equation*}%
where the upper sign corresponds to an incident clockwise circularly
polarized EM, and the lower sign to an anti-clockwise one. Let us define as
a shorthand notation%
\begin{equation}
z_{\pm }\equiv x\pm iy
\end{equation}%
as the complex representation of the displacement of the electron. Solving
for $z_{\pm }$, one obtains%
\begin{equation}
z_{\pm }=\frac{eE_{\pm }}{-m\left( \omega ^{2}\mp \omega \omega _{\text{cycl}%
}\right) }
\end{equation}%
where the cyclotron frequency $\omega _{\text{cycl}}$ is defined as%
\begin{equation}
\omega _{\text{cycl}}\equiv \frac{eB}{mc}\text{ ,}
\end{equation}%
and where%
\begin{equation*}
E_{\pm }\equiv E_{x}\pm iE_{y}\text{ .}
\end{equation*}%
For a gas of electrons with a uniform number density $n_{e}$, the
polarization of this medium induced by the weak incident EM wave is given by%
\begin{equation}
P_{\pm }=n_{e}e\left( x\pm iy\right) =n_{e}ez_{\pm }=\frac{n_{e}e^{2}E_{\pm }%
}{-m\left( \omega ^{2}\mp \omega \omega _{\text{cycl}}\right) }=\chi
_{e}E_{\pm }
\end{equation}%
where the susceptibility of the electron gas is given by%
\begin{equation}
\chi _{e}=\frac{n_{e}e^{2}}{-m\left( \omega ^{2}\mp \omega \omega _{\text{%
cycl}}\right) }=-\frac{\omega _{\text{plas}}^{2}/4\pi }{\omega ^{2}\mp
\omega \omega _{\text{cycl}}}
\end{equation}%
where the plasma frequency $\omega _{\text{plas}}$ is defined by%
\begin{equation}
\omega _{\text{plas}}\equiv \sqrt{\frac{4\pi n_{e}e^{2}}{m}}\text{ .}
\end{equation}%
The index of refraction of the gas $n(\omega )$\ is given by%
\begin{equation}
n(\omega )=\sqrt{1+4\pi \chi _{e}(\omega )}=\sqrt{1-\frac{\omega _{\text{plas%
}}^{2}}{\omega ^{2}\mp \omega \omega _{\text{cycl}}}}\text{ .}
\end{equation}%
Specular reflection occurs when the index of refraction becomes a pure
imaginary number. Let us define as the critical frequency $\omega _{\text{%
crit}}$ as the frequency at which the index vanishes, which occurs when%
\begin{equation}
\frac{\omega _{\text{plas}}^{2}}{\omega _{\text{crit}}^{2}\mp \omega _{\text{%
crit}}\omega _{\text{cycl}}}=1.
\end{equation}%
Since the index vanishes at this critical frequency, the Fresnel reflection
coefficient ${\mathcal{R}}(\omega )$ from the planar structure for normal
incidence at criticality is given by%
\begin{equation}
{\mathcal{R}}(\omega )=\left\vert \frac{n(\omega )-1}{n(\omega )+1}%
\right\vert ^{2}\rightarrow 100\%\text{ when }\omega \rightarrow \omega _{%
\text{crit}}\text{ ,}  \label{Fresnel-Reflection-for-plasma}
\end{equation}%
which implies specular reflection of the incident plane EM wave from the
electron gas. \ This yields a quadratic equation for $\omega _{\text{crit}}$%
\begin{equation}
\omega _{\text{crit}}^{2}\mp \omega _{\text{crit}}\omega _{\text{cycl}%
}-\omega _{\text{plas}}^{2}=0.
\end{equation}%
The solution for $\omega _{\text{crit}}$ is%
\begin{equation}
\omega _{\text{crit}}=\frac{\pm \omega _{\text{cycl}}\pm \sqrt{\omega _{%
\text{cycl}}^{2}+4\omega _{\text{plas}}^{2}}}{2}.  \label{quadratic-solution}
\end{equation}%
The first $\pm $ sign is physical, and is determined by the sense of
circular polarization of the incident plane wave. The second $\pm $ sign is
mathematical, and originates from the square root. \ One of the latter
mathematical signs is unphysical. To determine which choice of the latter
sign is physical and which is unphysical, let us first consider the limiting
case when the inequality%
\begin{equation}
\omega _{\text{cycl}}<<\omega _{\text{plas}}
\end{equation}%
holds. This inequality corresponds physically to the situation when the
magnetic field is very weak, but the electron density is very high, so that
the phenomenon of specular reflection of EM waves with frequencies below the
plasma frequency $\omega _{\text{plas}}$ occurs. Let us therefore take the
limit $\omega _{\text{cycl}}\rightarrow 0$ in the solution (\ref%
{quadratic-solution}). Negative frequencies are unphysical, so that we must
choose the positive sign in front of the surd as the only possible physical
solution. Thus in general it must the case that the physical root of the
quadratic is given by%
\begin{equation}
\omega _{\text{crit}}=\frac{\pm \omega _{\text{cycl}}+\sqrt{\omega _{\text{%
cycl}}^{2}+4\omega _{\text{plas}}^{2}}}{2}.  \label{Physical-root}
\end{equation}%
\qquad

\bigskip Let us now focus on the more interesting case where the magnetic
field is very strong, but the number density of electrons is very small, so
that the plasma frequency is very low, corresponding to the inequality 
\begin{equation}
\omega _{\text{cycl}}>>\omega _{\text{plas}}\text{ .}
\end{equation}%
There then are two possible solutions, corresponding to clockwise-polarized
and anti-clockwise-polarized EM waves, respectively, viz.%
\begin{equation}
\omega _{\text{crit,1}}=\omega _{\text{cycl}}\text{ and }\omega _{\text{%
crit,2}}=0\text{ .}  \label{2-physical-solutions}
\end{equation}%
Note the important fact that these solutions are independent of the number
density (or plasma frequency) of the electron gas, which implies that even a
very dilute electron gas system can give rise to specular reflection. \ The
fact that these solutions are independent of the number density also implies
that they would apply to the case of an inhomogeneous electron density, such
as that arising for a single delocalized electron confined to the vicinity
of the plane $z=0$ by the Coulombic attraction to its image. Both solutions
of the quadratic equation (\ref{2-physical-solutions}) are now physical
ones, and imply that whether the sense of rotation of the EM polarization
co-rotates or counter-rotates with respect to the magnetic-field--induced
precession of the guiding center motion of the electron around the magnetic
field, determines which sense of circular polarization is transmitted when $%
\omega >$ $\omega _{\text{crit,2}}=0$, or which sense of circular
polarization is totally reflected when $\omega <\omega _{\text{crit,1}%
}=\omega _{\text{cycl}}$, provided that the frequency of the incident
circularly polarized EM wave is less than the cyclotron frequency $\omega _{%
\text{cycl}}$. \ The interesting solution is the one with the non-vanishing
critical frequency, since it implies that there always exists one solution
where there is specular reflection of the EM wave, even when the number
density of electrons is extremely low, i.e., even when the plasma frequency $%
\omega _{\text{plas}}$ approaches zero, and even when this number density
becomes very inhomogeneous as a function of $z$.

In the extreme case of a single electron completely delocalized on the
outside surface of superfluid helium (for example, in an $S$ state on the
outside surface of a spherical drop), one should solve the problem quantum
mechanically, by going back to Landau's solution of the motion of an
electron in a uniform magnetic field, and adding as a time-dependent
perturbation the weak (classical) incident circularly polarized plane wave.
\ However, the above classical solution should hold in the correspondence
principle limit, where, for the single delocalized electron, the effective
number density of the above classical solution is determined by the absolute
square of the electron wavefunction, viz.%
\begin{equation}
n_{e}=\left\vert \psi _{e}\right\vert ^{2}\text{ , and}
\end{equation}%
\begin{equation}
\int n_{e}dV=\int \left\vert \psi _{e}\right\vert ^{2}dV=1.
\end{equation}%
Here we must take into account the fact that there is a finite confinement
distance $d_{e}\approx $ 80\AA\ in the $z$ direction of the electron's
motion in the hydrogenic ground state caused by the Coulomb attraction of
the electron to its image charge induced in the dielectric, but the electron
is completely delocalized in the $x$ and $y$ directions on an arbitrarily
large plane (and hence over the large spherical surface of a large drop).
The effective plasma frequency of the single electron may be extremely
small; nevertheless, total reflection by this single, delocalized electron
still occurs, provided that the frequency of the incident circularly
polarized EM wave is below the cyclotron frequency. The fundamental reason
why even just a single delocalized electron in a strong magnetic field can
give rise to specular reflection is that the $\mathbf{v\times B}$ Lorentz
force \cite{Gravito-Hall-effect} leads to a longitudinal quantum Hall
resistance that is strictly zero, which shorts out the incident circularly
polarized EM wave. Thus one concludes that the hard-wall boundary conditions
used in the order-of-magnitude estimate given by Equation (\ref%
{Geometric-X-section}) of the scattering cross-section of microwaves from
the drops are reasonable ones. This conclusion will be tested experimentally
(see Figure (\ref{Rescaled-scattering-experiment-4})).

\bigskip \textbf{Acknowledgments} I thank John Barrow, Fran\c{c}ois
Blanchette, George Ellis, Sai Ghosh, Dave Kelley, Tom Kibble, Steve Minter,
Kevin Mitchell, James Overduin, Richard Packard, Jay Sharping, Martin
Tajmar, Roland Winston, and Peter Yu for their help.


\begin{thebibliography}{99}
\bibitem{MTW} C. W. Misner, K. S. Thorne, and J. A. Wheeler, \textit{%
Gravitation} (Freeman, San Francisco, 1972): \textquotedblleft
MTW\textquotedblright\ henceforth.

\bibitem{Landau} L.~Landau and E.~Lifshitz \textit{The Classical Theory of
Fields}, 1st edition (Addison-Wesley, Reading, MA, 1951), page 331, Equation
(11-115).

\bibitem{Weinberg} S. Weinberg, \textit{Gravitation and Cosmology} (John
Wiley \& Sons, New York, 1972).

\bibitem{Taylor1994} J.~G.~Taylor, Rev.~Mod.~Phys.~\textbf{66}, 711 (1994).

\bibitem{Lamb-medal} R. Y. Chiao, J. Mod. Opt. \textbf{53}, 2349 (2006)
(quant-ph/0601193).

\bibitem{Wald} R. M. Wald, \textit{General Relativity} (University of
Chicago Press, Chicago, 1984). The Maxwell-like equations Equations (\ref%
{Maxwell-like-eq-1}) - (\ref{Maxwell-like-eq-4}) were derived by Wald in
section 4.4 by starting from the assumption that for weak gravitational
fields, the metric of spacetime can be approximated by (in the notation of
MTW \cite{MTW})%
\begin{equation}
g_{\mu \nu }\approx \eta _{\mu \nu }+h_{\mu \nu }
\end{equation}%
where $g_{\mu \nu }$ is the metric tensor, $\eta _{\mu \nu }$ is the
Minkowski metric tensor for a flat spacetime, and $h_{\mu \nu }$ are small
perturbations of the metric tensor, such as those arising from gravitational
radiation. When the lowest order effects of the motion of the source are
taken into account, but neglecting stresses, the linearized Einstein field
equations, when also linearized in the velocity of the matter, become (in
units where $G=c=1$)%
\begin{equation}
\partial ^{\mu }\partial _{\mu }\overline{h}_{0\lambda }=16\pi J_{\lambda }
\label{wave-eq-for-h_0i}
\end{equation}%
where $\overline{h}_{\mu \nu }=h_{\mu \nu }-\frac{1}{2}\eta _{\mu \nu }h$
and where $J_{\lambda }$ is the mass current density four-vector of the
source. If, following Wald, one defines the \textquotedblleft vector
potential\textquotedblright\ as follows:%
\begin{equation}
A_{\mu }\equiv -\frac{1}{4}\overline{h}_{\mu \nu }t^{\nu }\text{,}
\end{equation}%
where $t^{\nu }$ is the four-velocity of a test particle, one obtains%
\begin{equation}
\partial ^{\mu }\partial _{\mu }A_{\lambda }=-4\pi J_{\lambda }\text{ .}
\end{equation}%
Therefore these equations have precisely the form of Maxwell's equations in
the Lorentz gauge, with the consequence that the perturbations $\overline{h}%
_{0\lambda }$ propagate with precisely the speed of light $c$, and not at
the speed $c/2$. In contrast to this, using the PPN formalism, V. B.
Braginsky, C. M. Caves, and K. S. Thorne, Phys. Rev. D \textbf{15}, 2047
(1977), derived a set of Maxwell-like equations which yielded a speed of $%
c/2 $, and not the speed of light $c$, for time-varying perturbations of the
fields. This difference in speeds arises from the fact that the PPN
formalism describes the near-fields as seen by an observer close to the
source, but Wald's formalism describes the far-fields as seen by an observer
in an asymptotically flat spacetime far away from the source. In the
transverse-traceless gauge, one of the gauge conditions is%
\begin{equation}
h_{0\mu }\equiv 0\text{ .}
\end{equation}%
An incorrect conclusion drawn from this gauge condition is that only the
gravito-electric components given by the strains $h_{ij}$ of a gravitational
plane wave exist, and that no gravito-magnetic components of radiation
fields in the far field of sources can exist. In F. de Felice and C. J. S.
Clarke, \textit{Relativity on Curved Manifolds} (Cambridge University Press,
Cambridge, 1973), Chap. 9, the authors point out that the Riemann curvature
tensor for gravity waves propagating in a flat background can be separated
into \textquotedblleft electric\textquotedblright\ and \textquotedblleft
magnetic\textquotedblright\ parts, and that these Riemann curvature tensor
components satisfy tensor Maxwell-like equations. The wave speed that
follows from these equations is again precisely $c$. This gauge-invariant
way of characterizing gravitational radiation shows that the
\textquotedblleft electric\textquotedblright\ and \textquotedblleft
magnetic\textquotedblright\ components of the Riemann curvature tensor for a
monochromatic gravitational plane wave propagating in the vacuum are equal
in magnitude to each other in natural units. The earliest mention of
Maxwell-like equations for linearized general relativity was perhaps made by
R. L. Forward in Proc. IRE \textbf{49}, 892 (1961).

\bibitem{Kiefer-Weber} C.~Kiefer and C.~Weber, Annalen der Physik (Leipzig) 
\textbf{14}, 253 (2005).

\bibitem{Chiao2004} R.~Y.~Chiao, in \textit{Science and Ultimate Reality},
eds.~J.~D.~Barrow, P.~C.~W.~Davies, and C.~L.~Harper, Jr.~(Cambridge
University Press, Cambridge, 2004), page 254 (quant-ph/0303100); R. Y.
Chiao, W. J. Fitelson, and A. D. Speliotopoulos (gr-qc/0304026); R. Y. Chiao
and W. J. Fitelson (gr-qc/0303089), \textquotedblleft Time and Matter in the
Interaction between Gravity and Quantum Fluids: Are There Macroscopic
Quantum Transducers between Gravitational and Electromagnetic
Waves?\textquotedblright\ published in the proceedings of the `Time \&
Matter Conference' in Venice, Italy, 11-17 August 2002, eds. Ikaros Bigi and
Martin Faessler (World Scientific, Singapore, 2006), page 85.

\bibitem{Tinkham} M. Tinkham, \textit{Introduction to Superconductivity},
2nd edition (Dover Books on Physics, New York, 2004).

\bibitem{Prange} R. E. Prange and S. M. Girvin, \textit{The Quantum Hall
Effect}, 2nd edition (Springer Verlag, New York, 1990).

\bibitem{Hossenfelder} Recall the boundary conditions that follow from
Maxwell's equations for electromagnetism. Consider for simplicity a planar
boundary. The local normal component of the magnetic field must be
continuous across the boundary (this comes from the Maxwell equation $%
\mathbf{\nabla \cdot B}=0$ applied to a small pillbox that straddles the
boundary), and the local tangential component of the magnetic field must
have a discontinuous jump across the boundary due to arbitrarily thin
surface currents flowing at the boundary (this comes from the Maxwell
equation $\mathbf{\nabla \times B}=\mu _{0}\mathbf{J}$, where $\mathbf{J}$
is the electric current density, applied to a small rectangular loop that
straddles the boundary). Let us assume that Ohm's law holds for these thin
surface currents, and take the limit as the surface resistance vanishes. For
a superconductor, and also for a quantum Hall fluid moving frictionlessly on
the surface of superfluid helium, the surface resistance of the electrons on
the surface is strictly zero. This leads to specular reflection of EM waves
from the boundary (e.g., the reflection of microwaves with frequencies well
below the BCS gap frequency from superconducting thin films \cite{Tinkham}).
But each electron carries mass as well as charge with it when it moves.
Therefore a strictly zero surface resistance in the electrical sector
implies a strictly zero surface resistance in the gravitational sector. The
gravitational Maxwell-like equations lead to the same local normal and
tangential boundary conditions for the gravito-magnetic field in the
gravitational sector as the ones given above for the electromagnetic sector.
Thus the zero surface resistance of the surface electrons leads in principle
to the specular reflection of GR waves at frequencies well below the
relevant gap frequency from the boundary, in precisely the same way that the
zero surface resistance of the same electrons leads to the specular
reflection of the EM waves at frequencies well below the relevant gap
frequency from the same boundary. While it is true that most of the mass is
in the interior to a \textquotedblleft Millikan oil drop,\textquotedblright\
for the validity of the specular boundary conditions, it is the linear
response of the electrons on the surface of the drop to the gravitational
radiation fields that is crucial. See footnotes \cite{Tinkham2} and \cite%
{Gravito-Hall-effect} as to the detailed reasons why specular boundary
conditions should apply to the quantum Hall fluid on the surface of a
\textquotedblleft Millikan oil drop.\textquotedblright

\bibitem{Tinkham2} It may be objected that Equations (\ref%
{Reflection-from-vacuum-superconductor-interface}), (\ref{Fresnel-reflection}%
), and (\ref{Fresnel-Reflection-for-plasma}) are believed to apply only when
the thickness $d$ of a sample is large compared with the relevant
penetration depth $\ell _{P}$, whereas the opposite limit (appropriate for a
thin-film sample) is assumed here. (In the case of superconductors, the
penetration depth $\ell _{P}$ is the London penetration depth $\lambda _{L}$%
.) Contrary to this common belief, for the case when the film is thin
compared to the penetration depth, but when the penetration depth is much
less than the radiation wavelength, i.e., $d<<\ell _{P}$ but $\ell
_{P}<<\lambda $, where $\lambda $ is the free space wavelength, the
reflectivity is not of the order of $(d/\ell _{P})^{2}$, as one might
naively expect, but is much higher, and in fact approaches unity as $\lambda 
$ becomes infinite. See Equation (3.128) of Tinkham's book \cite{Tinkham}
for the transmissivity ${\mathcal{T}}$ of superconducting thin films, which
reads as follows:%
\begin{equation}
{\mathcal{T}}=\left[ \left( 1+\frac{\sigma _{1}Z_{0}d}{n+1}\right)
^{2}+\left( \frac{\sigma _{2}Z_{0}d}{n+1}\right) ^{2}\right] ^{-1}\text{ ,}
\label{Tinkham's-(2.128)}
\end{equation}%
where $\sigma =\sigma _{1}+i\sigma _{2}$ is the complex conductivity of the
thin film, $d$ is its thickness, $n$ is the index of refraction of its
substrate, and $Z_{0}=\sqrt{\mu _{0}/\varepsilon _{0}}=\mu _{0}c$ is the
characteristic impedance of free space for EM waves. Although this equation
was derived by Tinkham in the context of superconductivity, it applies to
all thin films with a complex conductivity $\sigma =\sigma _{1}+i\sigma _{2}$%
. (It can also be readily generalized to the case of a complex conductivity 
\emph{tensor} which is applicable to the quantum Hall fluid.) From this
equation, we see that the transmissivity can vanish in the low-frequency
limit $\omega \rightarrow 0$, since for superconductors, $\sigma
_{2}\rightarrow 1/\omega \rightarrow \infty $, leading to a substantial
reflection of these waves, there being a negligible dissipation within the
superconducting film. This result can be understood in terms of an
inductance per square element of the thin film%
\begin{equation}
L=\mu _{0}\ell _{\text{gap}}
\end{equation}%
where $\ell _{\text{gap}}$ is a characteristic gap length scale of the
superconductor or of the quantum Hall fluid. This leads to a reactance per
square element of the film of%
\begin{equation}
X_{L}=\omega L=\frac{1}{\sigma _{2}d}
\end{equation}%
whose low value is responsible for the high reflectivity for waves with
frequencies well below the relevant gap frequency. However, in the
derivation of Equation (3.128) of Tinkham's book, it was assumed that the
thin conducting film sample was transversely infinite, so that it is not
immediately obvious that it can be applied to the electrons on a spherical
\textquotedblleft Millikan oil drop,\textquotedblright\ nor is it clear that
the concept of a \textquotedblleft penetration depth\textquotedblright\
applies to the quantum Hall fluid on the surface of superfluid helium.
Nevertheless, the only relevant length scales for this fluid are the
magnetic length scale (in SI units) $\ell _{B}=(h/eB)^{1/2}$\ for the
quantum Hall effect, and the confinement distance scale $d_{e}$ of electrons
on the superfluid drop surface discussed in Appendix A, both of which are on
the order of 10 nm \cite{Prange}\cite{Grimes2}, whereas the radius of a
typical drop is around 4 mm, which is much larger than both of these
microscopic length scales. Since a small patch on the surface of a large
spherical drop looks planar on these length scales, one can still apply
locally to this small patch, in the limit of long wavelengths $\lambda $,
the discontinuous-jump boundary conditions (see footnote \cite{Hossenfelder}%
) for the tangential magnetic field that follows from the Maxwell equation $%
\mathbf{\nabla \times B}=\mu _{0}\mathbf{J}$ and from its gravitational
analog $\mathbf{\nabla \times B}_{G}=\mu _{G}\mathbf{J}_{G}$. It is these
discontinuous-jump boundary conditions for the tangential components of both 
$\mathbf{B}$ and $\mathbf{B}_{G}$ that leads to nonnegligible reflections of
both EM and GR waves from the quantum Hall fluid on the surface of a drop.
They are also the basis for Equation (3.128) of Tinkham's book. Therefore
the planar model used in the derivation of Equation (3.128) of Tinkham's
book\ should be valid for the reflectivity of the spherical
\textquotedblleft Millikan oil drops\textquotedblright\ being considered for
the proposed experiment.\ See footnote \cite{Gravito-Hall-effect} for a
discussion of the physical origin of the surface currents responsible for
the reflection in the case of GR waves. In the case of EM waves, the
transmissivity of EM waves at low frequencies is given by%
\begin{equation}
{\mathcal{T}}\approx 4\left( \frac{\omega L}{Z_{0}}\right) ^{2}=4\left( 
\frac{\omega \mu _{0}\ell _{\text{gap}}}{\mu _{0}c}\right) ^{2}=4\left( 
\frac{2\pi \ell _{\text{gap}}}{\lambda }\right) ^{2}
\label{Transmissivity-in-low-frequency-limit}
\end{equation}%
where the approximation has been made that $n\approx 1$. Thus ${\mathcal{T}}$%
\ is on the order of $(\ell _{\text{gap}}/\lambda )^{2}=(\omega /\omega _{%
\text{gap}})^{2}\approx (\omega /\omega _{\text{cycl}})^{2}$, since $\omega
_{\text{gap}}\approx \omega _{\text{cycl}}$ in the case of the quantum Hall
fluid. (See Appendix A.) Thus the transmission ${\mathcal{T}}$ both of a
superconducting thin film and of a quantum Hall fluid film remains small,
and therefore the reflectivity $\mathcal{R}=1-{\mathcal{T}}$\ of these films
remains high for all frequencies $\omega $\ of an incident wave which are
well below the relevant gap frequency $\omega _{\text{gap}}$. Note that the
permeability of free space $\mu _{0}$ cancels out of Equation (\ref%
{Transmissivity-in-low-frequency-limit}), and therefore that $\mu _{G}$ will
also cancel out of the analogous expression for the case of GW waves.
Therefore, since the quantum Hall fluid is strictly dissipationless, there
results a nonnegligible reflectivity for both EM and GR waves from the
\textquotedblleft Millikan oil drops\textquotedblright\ for waves with
frequencies well below the relevant gap frequency, i.e., the cyclotron
frequency $\omega _{\text{cycl}}$.\ Now we turn from the case of quantum
Hall fluids to that of superconductors. In connection with Equation (\ref%
{Fresnel-reflection}), it is commonly believed that the gravitational analog
of the London penetration depth of a superconductor is many orders of
magnitude larger than $\lambda $, so that it would seem that Equation
(3.128) of Tinkham's book cannot be applied to superconductors in the
gravitational sector. However, two points need to be made in this regard.
First, the concept of a \textquotedblleft gravitational analog of the London
penetration depth\textquotedblright\ may not apply to superconductors in the
first place, due to the anti-Meissner effect (see the discussion following
Equation (\ref{continuity})). The Yukawa equation for the electromagnetic
London penetration depth for a superconductor must be replaced by the
Helmholtz equation in order to describe the behavior of the gravito-magnetic
field $\mathbf{B}_{G}$ inside the superconductor. Second, the large value of
the ferromagnetic-like enhancement factor $\kappa _{G}^{\text{(magn)}}$ must
be taken properly into account in the numerical value of the
\textquotedblleft gravitational analog of the London penetration
depth,\textquotedblright\ if Equation (\ref{mu-G-prime}) is indeed the
correct explanation for Tajmar's experiment \cite{Tajmar}. I thank an
anonymous referee for pointing out to me Equation (3.128) of Tinkham's book 
\cite{Tinkham}.

\bibitem{Tajmar} M. Tajmar, F. Plesescu, B. Seifert, and K. Marhold,
preprint gr--qc/0610015; M. Tajmar, F. Plesescu, K. Marhold, and C. J. de
Matos, pre-print gr-qc/0603034, (2006); C. J. de Matos, and M. Tajmar,
Physica C \textbf{432}, 167 (2005).

\bibitem{Tajmar2} Martin Tajmar (private communication).

\bibitem{Kramers-Kronig} The response of the medium must not only be \emph{%
linear} in the amplitude of the weak applied gravitational radiation fields,
but it must also be \emph{causal}. \ Hence the real and imaginary parts of
the linear response function $\kappa _{G}^{\text{(magn)}}(\omega )$, as a
function of frequency $\omega $ of the gravity wave, must obey
Kramers-Kronig relations similar to those given by Equations (4) and (5) of
Ref. \cite{Chiao2004}(a).

\bibitem{Weilert1996} M.~A.~Weilert, D.~L.~Whitaker, H.~J.~Maris, and
G.~M.~Seidel, Phys.~Rev.~Letters \textbf{77}, 4840 (1996); M.~A.~Weilert,
D.~L.~Whitaker, H.~J.~Maris, and G.~M.~Seidel, J. Low Temp. Phys. \textbf{106%
}, 101 (1997). Another method of levitating charged superfluid drops uses a
parallel-plate capacitor geometry similar to that used by Millikan in his
original oil-drop experiment; see J. J. Niemela, J. Low Temp. Phys. \textbf{%
109}, 709 (1997).

\bibitem{Grimes2} C. C. Grimes and T. R. Brown, Phys. Rev. Letters \textbf{32%
}, 280 (1974); C. C. Grimes and G. Adams, Phys. Rev. Letters \textbf{36},
145 (1976). In the ground state of the system, the electron resides on the 
\emph{outside} surface of a superfluid helium drop, and not within the \emph{%
inside} volume of the drop. When the electron is forced to be within the
interior of the drop, it will form a bubble with a radius of around one
nanometer, due to the balancing of an outwards Pauli pressure with the
surface tension of the superfluid (see R. J. Donnelly, \textit{Experimental
Superfluidity} (The University of Chicago Press, Chicago, 1967), p. 176 ff).
The bubble will then rise to the surface, driven by the Coulomb force of
attraction to its own image charge induced in the surface. It will then
burst through the surface to uniformly coat the drop with one electron
charge on its outside surface. The electron will be in an $S$-state in order
to minimize the energy of the system. This then is the ground state of the
system.

\bibitem{Footnote-A} Note that the quantum-mechanical ground-state
wavefunction (or complex order parameter) must remain \textit{single-valued}
(according to a distant inertial observer) globally at all times everywhere
inside the interior of the system during the passage of a gravity wave. This
is another aspect of the \textquotedblleft quantum
rigidity\textquotedblright\ of a quantum fluid in its response to the
gravity wave.

\bibitem{Yu} Time-reversal symmetry under the \emph{global} operation of
time reversal includes here the reversal of the direction of the applied DC
magnetic fields.

\bibitem{Yariv1967} A.~Yariv, \textit{Quantum Electronics}, 1st edition
(John Wiley \& Sons, New York, 1967), page 223ff.{}

\bibitem{Gigantic-atom} This single quantum entity can be viewed as if it
were a gigantic atom, in which the usual atomic nucleus is replaced by the
superfluid helium drop, and the usual electronic cloud surrounding the
atomic nucleus is replaced by the electrons on the surface surrounding the
drop. The large energy gap (Equation (\ref{Cyclotron-gap})) arising from the
large applied magnetic field is what makes this gigantic atom extremely
rigid and dissipationless at low temperatures. A pair of such gigantic atoms
forms a gigantic diatomic molecule. If the charges and masses of the two
drops are slightly different from each other, such a gigantic diatomic
molecule may form an entangled state of charge and mass in its ground state,
provided that decoherence does not occur during the observation time. 

\bibitem{Laughlin} R. B. Laughlin, Phys. Rev. Letters \textbf{50}, 1395
(1983).

\bibitem{Footnote-B} The principle of equivalence should apply to all
charges and fields in curved spacetime \cite{Landau}\cite{Lamb-medal}.
However, Maxwell's equations, as usually formulated for standard
electromagnetism, are expressed in terms of fields on a \emph{flat}
spacetime. They must be generalized to fields on a \emph{curved} spacetime
when interactions with gravitational radiation are considered. The
back-action of EM waves propagating in a curved spacetime upon GR waves can
in principle arise from the contribution of the Maxwell stress-energy
tensor, which is \emph{quadratic} in the EM field strengths, as a source
term on the right-hand side of Einstein's field equations. Such quadratic
terms would give rise to second harmonic generation in the conversion of EM
to GR waves, but not to first harmonic generation. However, there can in
principle arise a \emph{linear} coupling of EM to GR waves when a strong DC
magnetic field is present, and Einstein's equations are linearized in the
weak EM and GR wave amplitudes. This linear coupling arises from a mixing
term, which consists of a product of the DC magnetic field strength and the
EM wave amplitude in the quadratic Maxwell stress-energy tensor that leads
to first harmonic generation of GR waves at the same frequency as that of
the incident EM waves in a linear scattering process. The role of the
\textquotedblleft Millikan oil drops\textquotedblright\ is that they can
greatly enhance the coupling between EM and GR waves due to their extreme
rigidity and large masses. The electrons on their surfaces tightly tie the
local \textbf{B} field lines to these drops, so that these lines are firmly
anchored to the drops. At very low temperatures when the system remains
adiabatically in the ground state, the \textbf{B} field lines and the drops
co-move rigidly together according to a distant observer, when the system is
disturbed by the passage of a GR or an EM wave. A given drop, however,
remains at rest with respect to a local inertial observer at the center of
the drop, and the local \textbf{B} field lines also do not appear to move
with respect to this local inertial observer. By contrast, to the distant
inertial observer in an asymptotically flat region of spacetime far away
from the drops, where radiation fields become asymptotically well defined,
the pair of drops appear to be in relative motion, and the system emits
power in both GR and EM radiations. Thus a graviton (spin 2) can in
principle be produced from a photon (spin 1) in the presence of a DC
magnetic field (spin 1), in a scattering process from the drops. See L.
Halpern, Arkiv f\"{o}r Fysik \textbf{35}, 57 (1967) for a
quantum-field-theoretic treatment of similar scattering processes. I thank
Tom Kibble for raising the important question: \textquotedblleft How can a
graviton (spin 2) be produced from a photon (spin 1)?\textquotedblright .

\bibitem{NASA2006} R. Y. Chiao in the proceedings of the NASA conference
\textquotedblleft Quantum to Cosmos\textquotedblright\ (quant-ph/0606118),
to be published in Int. J. Mod. Phys. D.

\bibitem{Gravito-Hall-effect} The $\mathbf{v\times B}$ Lorentz force also
leads to a gravito-Hall effect, in which an electron, when subjected to a
gravitational field $\mathbf{g}$ in a dissipationless sample, moves with an
average velocity which is perpendicular to both $\mathbf{g}$ and $\mathbf{B}$
fields. For example, an electron in a vertically oriented, planar quantum
Hall sample subjected to the Earth's gravity field will move with an average
velocity at right angles to both the Earth's $\mathbf{g}$ field and the DC
magnetic $\mathbf{B}$ field applied normally to the sample. This then
induces a Hall current which is directly proportional to, and perpendicular
to, the applied $\mathbf{g}$ field. Local, time-varying gravitational fields 
$\mathbf{g}(t)$ (as seen by a distant observer) arising from gravitational
radiation impinging at normal incidence to the sample, will induce
time-varying transverse electrical currents in the quantum Hall sample in
the presence of a strong DC magnetic field. Since each electron carries mass
as well as charge with it when it moves, this radiation will also induce
transverse, time-varying mass currents in this sample. The analysis in
Appendix A can be generalized to include the case of incident gravitational
radiation, once the quadrupolar nature of this radiation is taken into
account. For one sense of circular polarization, a 180$^{\circ }$ phase
shift between the transmitted and incident radiation fields leads to the
destructive interference of the transmitted and incident radiation fields,
independent of whether these fields are EM or GR in nature. The destructive
interference of the transmitted wave with the incident wave in the forwards
direction leads to reflection of the incident wave in the backwards
direction. Thus the longitudinal quantum Hall resistance in both EM and GR
sectors vanishes due to \textquotedblleft quantum
dissipationlessness,\textquotedblright\ so that circularly polarized EM and
GR radiation fields of one sense are both shorted out, leading to the
specular reflection of both kinds of incident radiations.
\end{thebibliography}
\end{document}